\documentclass[preprint2]{aastex}
\usepackage{t1enc}
\usepackage{graphicx}
\usepackage{epstopdf}
\usepackage{natbib}
\usepackage{lscape}
\usepackage{subfigure}
\usepackage{mathtools}
\usepackage{float}
\usepackage[flushleft]{threeparttable}
\usepackage{lineno}
\usepackage{acronym}
\usepackage{longtable}

\newcommand{\gap}%
{\raisebox{-0.5ex}{$\stackrel{\scriptstyle >}{\scriptstyle \sim}$}}

\DeclareUnicodeCharacter{2212}{-}

\begin{document}

\shorttitle{ The Population of compact radio sources in M\,17}
\shortauthors{Yanza et al.}

\title{THE POPULATION OF COMPACT RADIO SOURCES IN M\,17}

\author{
Vanessa Yanza\altaffilmark{1,2},
Josep M. Masqu\'e\altaffilmark{1},
Sergio A. Dzib\altaffilmark{3},
Luis F. Rodr\'iguez\altaffilmark{2},
S.-N.X. Medina\altaffilmark{3},
Stan Kurtz\altaffilmark{2},
Laurent Loinard\altaffilmark{2}, 
Miguel A. Trinidad\altaffilmark{1},
Karl M. Menten\altaffilmark{3},
Carlos A. Rodr\'iguez-Rico\altaffilmark{1} 
} 

\altaffiltext{1}{Departamento de Astronom\'ia, Universidad de Guanajuato, Apdo. Postal 144, 36000 Guanajuato, M\'exico}
\altaffiltext{2}{Instituto de Radioastronom\'ia y Astrof\'isica, Universidad Nacional Aut\'onoma de M\'exico, Morelia 58089, M\'exico}
\altaffiltext{3}{Max Planck Institut f\"ur Radioastronomie, Auf dem H\"ugel 69, D-53121 Bonn, Germany}


\begin{abstract}

We present a catalog of radio sources of the M\,17 region based on deep X band radio observations centered at 10 GHz obtained with the Jansky Very Large Array in the A configuration. We detect a total of 194 radio sources, 12 of them extended and 182 compact. We find that a significant fraction (at least 40\% in our catalog) have suspected gyrosynchrotron emission associated with stellar coronal emission. By comparing the radio luminosities of our sources with their X ray counterparts, when available, we find that they are underluminous in X rays with respect to the G\"udel Benz relation, but a correlation with a similar slope is obtained provided that only sources with evident non thermal nature are selected from the sample compiled for the Orion Nebula Cluster (ONC) and M\,17. The comparison of M\,17 with the ONC and NGC\,6334D-F leads to a similar luminosity function for the three regions, at least for the more luminous sources. However, the radio sources in M\,17 are three times more numerous compared to the other regions at a given luminosity and their spatial distribution differs from that of Orion. Moreover, an arc-shaped structure of 40$\arcsec$ in extent is observed in our map, identified previously as an ionizing front, with the cometary Hyper Compact source UC1 at its focus. Archival 1 mm ALMA data reveals compact emission coincident with the radio wavelength peak, possibly associated with a protostellar disk of the massive star exciting UC1.  

\end{abstract}


\section{Introduction}

The Messier 17 (M\,17) region (the Omega Nebula, Horseshoe Nebula, Swan Nebula, Checkmark Nebula, W\,38, S\,45 or NGC\,6618) is a bright HII region associated with a giant molecular cloud located in the Sagittarius constellation at a distance of $1.98_{-0.12}^{+0.14}$~kpc \citep{wu2014}. The complex is considered one of the most massive and luminous regions of recent star formation in the Milky Way \citep{lada1974}. It is composed of a main star-forming region located between two bars situated to the North (N-bar) and to the South (S-bar, Felli et al. 1980).  This region has an associated open stellar cluster \citep{beetz1976,morales2013} whose most luminous object is the massive multiple system CEN 1 \citep{chini1980}. The main components of this system, CEN\,1a and CEN\,1b \citep{hoffmeister2008} with nearly equal masses, have been detected at several wavelengths and found to be time variable in some of them \citep{broos2007,rodriguez2009}. 

The cluster ionizes the surrounding interstellar medium (ISM) so that a large arc-shaped  structure is formed \citep{felli1980,johnson1998} that is associated with an ionization front. Beyond this limit the material of the cloud is mostly molecular \citep{felli1980,felli1984}. At the focus of the arc structure there is a very radio-bright Hyper-Compact HII (HCHII) region called UC1 \citep{felli1980}. Many studies dedicated to this source show that it is associated with 
a disk-like structure and masers \citep{johnson1998, nielbock2007} among other features. Given the peculiar location of UC1 with respect to the ionization front, \citet{felli1984} and \citet{johnson1998} discuss the possibility of shock-induced star formation occurring in this source.  

The large luminosity of M\,17 ($\sim10^7~L_\odot$) and its extended size  \citep[up to 1\arcdeg\ considering the molecular environment, see][]{elmegreen1977a}, together with the wealth of structures found within, motivated many surveys of the region at multiple wavelengths. At IR and optical bands, following the early survey of \citet{chini1980}, the population of massive and intermediate mass young objects came to light in the subsequent surveys \citep{chini1985,chini1998, hanson1997,nielbock2007}. Moreover, deeper observations show that beyond the central cluster of massive stars, additional generations of young stellar objects populating other parts of the complex are present \citep{jiang2002, chini2004, povich2009}. This stellar richness was confirmed by \citet{broos2007} with X-ray observations, which detected 886 sources around M\,17 covering a wide field. At radio wavelengths, \citet[][hereafter RGM2012]{rodriguez2012} detected 38 compact radio sources, 19 of them with stellar counterparts. These 38 sources constitute the most extensive catalog of radio sources in M\,17 to date.  

\defcitealias{rodriguez2012}{RGM2012}

Because star forming regions are embedded in dense molecular clouds, radio emission counterparts provide information on the most obscured objects associated to the region. Furthermore, since the whole region is a strong radio emitter due to the extended gas ionized by massive stars, interferometric observations are crucial to isolate and explore the most compact parts. Such observations reveal the presence of a rich variety of types of radio sources, some of them individual and compact, suggesting that they are associated with young objects embedded in the parental nebula; many of them are not detected at any other wavelength.

The radio emission of these Compact Radio Sources (CRSs) can be thermal and/or non-thermal. The former is produced by free-free emission from ionized gas present in Hyper-Compact (HC)HII regions surrounding OB stars, ionized winds of massive stars, externally ionized protoplanetary disks (proplyds) or jets from low-mass protostars  \citep{panagia1975,shull1980,avalos2009, sanchez-monge2011b, zapata2004, stecklum1998, reynolds1986, anglada2018}. On the other hand, non-thermal radio emission is generated by relativistic charged particles moving in a magnetic field. This emission can be produced by gyrosynchrotron processes occurring in low-mass stars with superficial magnetic fields, or synchrotron processes due to the collision between winds of massive binary stars, among other possibilities \citep{feigelson1999, dzib2015, dougherty2000, blomme2013}.

The extended family of possible radio emitting objects described above implies a wide range of radio flux density expected for them. As a consequence, weak sources found at the lower end of this range become potentially undetectable and surveys might be biased to strong sources \citep{medina2019, brunthaler2021}. This has been alleviated in recent years thanks to the upgraded instrumentation installed on radio interferometers. This enables deep studies of star forming regions, that have notably increased the number of detected CRSs. Using one of the most important instruments, the  Very Large Array (VLA), several studies were conducted of different HII regions and very faint CRSs were detected. For example, early surveys of the Orion Nebula reported $\sim 20$ radio sources of a few mJy of flux density \citep{churchwell1987, garay1987}. \citet{zapata2004} later increased the cataloged sources to $77$ and found that a significant fraction of CRSs show time variability. Recently, \citet{forbrich2016} presented the deepest observations ever performed on Orion using the same instrument, upgraded to the {\it Karl G. Jasky} Very Large Array and detected $556$ radio sources of differing nature. The Nebula NGC\,6334, a string of star forming regions of different evolutionary status, including several HII regions, was studied recently by \citet{medina2018} who cataloged $83$ CRSs; only a few were previously known. Despite the larger distances of other regions as compared to Orion, in general, sufficiently deep observations of star forming regions show a significant increase in radio source detections compared to the old catalogs. This suggests that a rich population of radio sources exists in many regions of star formation; in most cases, this rich population remains to be uncovered. \\

Following previous radio observations of the M\,17 region \citepalias{rodriguez2012}, we carried out a study toward this region with the upgraded VLA. We achieve a significantly better sensitivity in our maps than in previous observations. This allows us to very significantly expand the radio source catalog of the M\,17 region by exploring the weakest radio components yet detected in the region. Furthermore, we analyze the nature of the population of radio sources using methods developed in other regions. In Section \ref{2}, we describe the observations and present the final map after the self-calibration process. In Section \ref{3}, we present the catalog of CRSs extracted from the final map. An analysis of the properties of CRSs extracted from the map is presented in Section \ref{4}. In Section \ref{5}, we constrain the nature of the CRSs and discuss their implications for the region. Finally, in Section \ref{6} we draw our conclusions. \\

\section{Observations} \label{2}

The VLA data were taken using the most extended A configuration in the X band (8-12 GHz) toward the M\,17 star forming region. The pointing center was $\alpha$(J2000) = 18$^{\mathrm{h}}$20$^{\mathrm{m}}$30$^{\mathrm{s}}$ and $\delta$(J2000) = $-16^\circ10'45''$  and the Full Width Half Maximum (FWHM) of the primary beam at the center of the band was 5$\rlap{$'$}$ (though it varies significantly over the 4 GHz bandwidth). The observations were made in three epochs with a total observation time of four hours (one hour on 2018 May 3, one hour on May 4 and two hours on May 8). The data were taken using the 3-bit sampling mode with an integration time set to 2 seconds. The correlator was configured to 4 GHz of continuous bandwidth coverage in full polarization mode, divided into 32 contiguous spectral windows, each with 64 channels of 2 MHz. The quasar 3C 286 was used as flux reference and bandpass calibrator, and J1832-1032 was used for the gain calibration. The theoretical noise of the combined observations is $2.6~\mu$Jy~beam$^{-1}$ and the expected synthesized beam is $\sim0\rlap{.}''2$.

\begin{figure}
    \centering
    \includegraphics[height=0.8\textwidth]{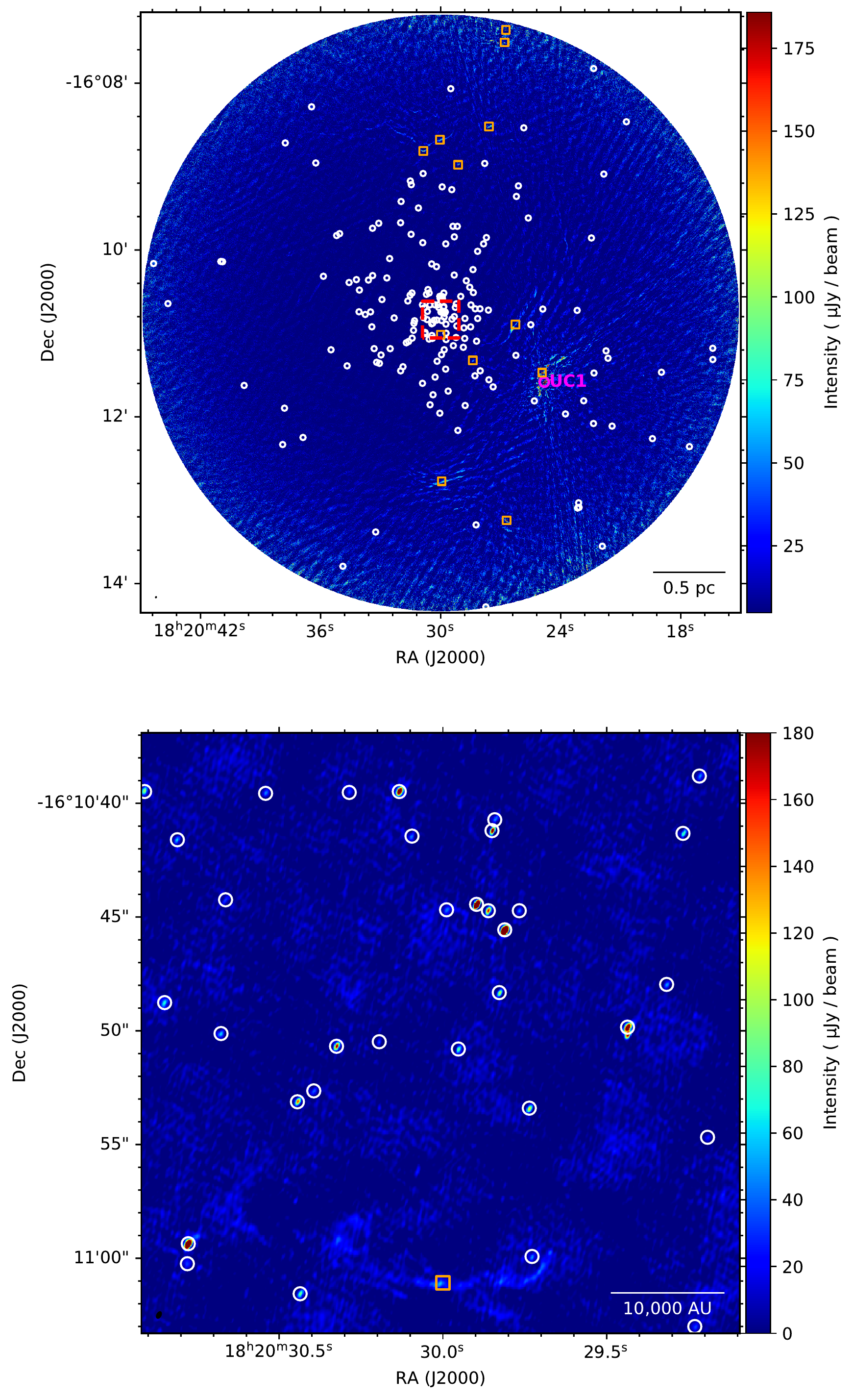}
\caption{\emph{Top panel:} VLA radio continuum map of the M\,17 region (color scale) using the A configuration and X band ($8-12$ GHz). The noise at its center is $5~\mu$Jy~beam$^{-1}$. The synthesized beam is $0\rlap{.}''285\times0\rlap{.}''152$, P.A. = $-28\rlap{.}^{\circ}77$. White circles represent the position of the compact radio sources detected. Orange squares reveal the position of the peak of the extended sources. The magenta circle represents the position of the brightest source, UC1. The increased noise at the outer edge of the image is due to the correction for the primary beam response. There is also increased noise in the surroundings of UC1. The dashed red box represents the area shown enlarged in the bottom panel. The full FITS image is available upon request to jmasque@ugto.mx. \emph{Bottom panel:} Zoom-in image of the central area of the observed field. Symbols are the same as in the top panel. The beam is the same as in the top panel and shown in the bottom left corner.
    \label{fig:bigim}}
\end{figure}

The data were edited, calibrated and imaged using the Common Astronomy Software Applications package, CASA \citep{McMullin2007}. An initial calibration and basic flagging were done following standard procedures. We combined the data of the three epochs into a single initial map using a weighting of {\it robust}$= - 0.5$. This provided a map with a beam size of $\sim0\rlap{.}''2$ and an rms noise of $27~\mu {\rm Jy~beam}^{-1}$. As this noise level was still far from the expected theoretical noise, we self-calibrated the combined data using the brightest source in the region, the UC1 source, assuming that this source does not vary significantly over the three observing days (\citetalias{rodriguez2012}). In this process we applied two calibrations in phase, followed by one in amplitude and, finally, one more in phase. This improved the noise level of the map to $5~\mu$Jy~beam$^{-1}$. This final map, which was primary beam-corrected, is shown in Fig.~\ref{fig:bigim}. Although the different primary beam function between the 9 and 11 GHz maps must bias the source detection toward sources with negative spectral indices, this effect is specially notable toward the outskirts of the map. Therefore, we adopted the FWHM of the primary beam at 9 GHz as the limiting size of the region where the spectral indices are trustable, which includes the majority of our sources. Moreover, most sources are gathered around the map center (i.e. in the HII region) where this effect is probably not significant.

\section{Results} \label{3}

\subsection{Source Extraction}

We extracted sources from the final map using the BLOBCAT software \citep{hales2012}. This software has been successfully used to do
source extractions from radio images of other star forming regions \citep[e.g.,][]{medina2018} and from radio surveys covering significantly larger areas of the sky
\citep[e.g., ][]{bihr2016,medina2019}. We follow a similar method to those used in these previous works which we briefly describe below.

BLOBCAT extracts information of islands of pixels using the flood fill algorithm in 2D astronomical images. It applies bias corrections to Gaussian and non-Gaussian sources to obtain accurate measurements. This allows us to perform an automatic extraction of radio sources and determine some of their observational properties. To correctly run the extraction, the Graphical Astronomy and Imaging Analysis Tool \citep[GAIA, based on SEXtractor;][]{bertin1996,holwerda2005} software was used to generate an $rms$ noise 2D map of the data. To obtain the background noise map, the chosen Photometry type and Checkimage were ``Background RMS'' and ``Full res. noise background'', respectively with a minimum island size of 5 pixels and a mesh size of 70 \citep[see][]{hales2012}. This noise map was then used with a signal to noise ratio of 5 (i.e., setting the $dSNR$ parameter of BLOBCAT to $5\sigma$) for the processing over our M\,17 map. With these parameters BLOBCAT extracts 200 preliminary radio sources.

\subsection{Extragalactic sources and false detections}

The expected number of background extragalactic sources in the imaged area was obtained using the equation below taken from \citet{anglada1998}. In this expression, $\theta_F$ is the angular diameter of the observed field, $\nu$ is the central frequency and $S_o$ is the minimum detected flux density at the field center. These values were set to 7.2$'$, 10 GHz, and 0.025 mJy, respectively.

\begin{equation}
\footnotesize
\centering
\label{eq:extrag}
\begin{split}
    \langle N \rangle ~ = 1.4~ \left\{ 1 - exp \left[ -0.0066 \left(\frac{\theta_F}{arcmin}\right)^2\left(\frac{\nu}{5~GHz}\right)^2 \right] \right\} \\ \left(\frac{S_o}{mJy}\right)^{-0.75} \left(\frac{\nu}{5~GHz}\right)^{-2.52} 
\end{split}
\end{equation}

This equation gives the expected number of extragalactic sources of $\langle N \rangle = 3 \pm 2$. In addition, using the statistical equation 1 from \citet{medina2018}, the maximum expected number of false detections is $5$. This suggests that a maximum of about 10 sources from the BLOBCAT outcome are artifacts or extragalactic sources. In consequence, almost all of the sources we catalog are real and potentially belong to the M\,17 region.


\subsection{Final catalog}

We visually inspected the BLOBCAT detections, defining them as compact or extended sources. We classify as extended sources those that have an irregular resolved morphology, and as compact those with an approximately Gaussian shape and a single well-defined peak.
Typical sizes for compact radio sources (CRSs) are not larger than a fraction of an arcsec. In some cases BLOBCAT detected several contiguous sources that are part of a single large structure. Specifically, 4 extended sources were comprised of 9 sources extracted by BLOBCAT. For these extended sources we re-measured the flux density using CASA tools, considering each one as a single source.  Moreover, we rejected 2 likely spurious sources by {\b visual} inspection, e.g. sources at the edge of the map where the {\it rms} noise level is high enough to create unreal sources. As a result, we retained 12 extended sources and 182 compact radio sources. These results are summarized in Table \ref{tab:selection}.

\begin{table*}[]
    \centering
    \caption{Summary of the source selection process to obtain the final catalog.}
    \begin{tabular}{c   c}
    \hline \hline
         Description & Number of sources \\ \hline
-         Total of sources extracted with BLOBCAT & 200 \\
          Rejected BLOBCAT sources\tablenotemark{a} & 9 \\
          Extended sources  & 8 \\
          Manually added extended sources & 4 \\
          Rejected BLOBCAT CRSs & 2 \\
          Manually added CRSs & 1 \\
         \hline
          Number of total extended sources & 12 \\
          Number of CRSs in the final catalog & 182 \\
         \hline
    \end{tabular}
    \label{tab:selection}
    \tablenotetext{a}{We interpreted these sources as parts of extended ones that are counted in the 4th row (see text).}
\end{table*}

Table \ref{tab:ext} shows the properties of the extended sources, namely their position (centered in the source emitting area), the peak flux, integrated flux density and angular area of each source measured by IMSTAT from CASA using contours of $5\sigma$. Sources 6 and 10 have been detected before by \citetalias{rodriguez2012} with their C band observations and are labeled 11 and 18 in their catalog, respectively. Their reported sizes are $0\rlap{.}''4$ and $0\rlap{.}''6$. Our measured sizes for these sources are approximately $1\rlap{.}''3 \times 0 \rlap{.}''4$ and $3\rlap{.}''7\times 2\rlap{.}''2$. This difference is likely due to the different sensitivity between the observations.


\begin{table*}
\caption{Extended sources in the M\,17 region from the final map.} \vspace{2mm}
\centering
\begin{tabular}{c	c	c	c	c c}
\hline
\hline
ID&{$\alpha$}[$^{\rm h}\,^{\rm m}\,^{\rm s}$] &{$\delta$}[$^\circ\,'\,''$]&\textbf{$S_{\rm peak}$}&\textbf{$S_{\rm int}$} & Area\\ 
\#&(J2000)&(J2000)&($\mu {\rm Jy~beam}^{-1}$)&($\mu$Jy)&(arcsec$^{2}$)\\  
\hline
1 & 18 20  24.931 & --16 11  28.29 & 560 $\pm$ 30 & 101425 $\pm$ 5433 & 40.3\\ 
2 & 18 20  29.950 & --16 12  46.44 & 117 $\pm$ 19 & 12341 $\pm$ 2004  & 20.5\\ 
3 & 18 20  26.264 & --16 10  53.71 & 158 $\pm$ 45 & 25439 $\pm$ 7245  & 18.9\\ 
4 & 18 20  30.042 & --16 08  40.72 &  76 $\pm$ 15 & 12770 $\pm$ 639 & 6.5\\
5 & 18 20  30.869 & --16 08  48.88 & 160 $\pm$ 15 & 9770 $\pm$ 489 & 6.1\\ 
6 & 18 20  27.588 & --16 08  31.24 & 103 $\pm$ 18 & 8533 $\pm$ 427 & 3.9 \\ 
7 & 18 20  26.698 & --16 13  14.56 & 244 $\pm$ 25 & 9800 $\pm$ 490  & 3.7\\
8 & 18 20  29.128 & --16 08  58.76 & 251 $\pm$ 16 & 4837 $\pm$ 242 & 3.4 \\ 
9 & 18 20  28.387 & --16 11  19.52 &  81 $\pm$ 10 & 2448 $\pm$ 123 & 2.2\\ 
10 & 18 20  30.008 & --16 11  01.08 & 57 $\pm$ 7 & 887 $\pm$ 109 & 2.1 \\ 
11 & 18 20  26.736 & --16 07  21.72 & 309 $\pm$ 43 & 1707 $\pm$ 237  & 0.5 \\ 
12 & 18 20  26.813 & --16 07  30.76 & 231 $\pm$ 37 & 1903 $\pm$ 304 & 0.4\\

\hline 
\end{tabular}
\label{tab:ext}
\end{table*}

\subsection{Counterparts at other wavelengths}

We searched for counterparts at other wavelengths cross-matching our CRSs using the SIMBAD database. The maximum discrepancy tolerance in position between our CRSs and the X-ray counterparts was $0\rlap{.}''8$.  For IR, optical and radio wavelengths it was $0\rlap{.}''75$. These limits include the astrometric uncertainty our VLA map and our fit, and the approximate position errors in the counterpart. In Table \ref{tab:CRS} (see Appendix A) we indicate the counterparts of our CRSs and their respective detected band.\\

We searched the Gaia catalog for counterparts to the radio sources identified in our survey. Table 3 lists the sources with Gaia counterparts, their parallax and distances. The errors on the distances were calculated using standard error propagation. We find that six of the sources have distances discrepant from that of M\,17 (d = 1.98 kpc) by more than two sigmas. These are sources with number 8, 27, 40, 110, 128 and 170. These sources are likely to be foreground objects, rather than sources associated with M\,17 itself. In the following, we exclude these sources our analysis. \\
  
\begin{table}[h!]
    \centering
    \caption{Distances of the Gaia counterparts of the CRSs}
    \begin{tabular}{c c c}
    \hline \hline
        ID & Parallax & Distance \\
        \# &	(mas) &	(kpc) \\
        \hline
6 &	0.57 $\pm$ 0.32 &	1.75 $\pm$ 0.98 \\
8 & 1.01 $\pm$ 0.19 & 0.99 $\pm$ 0.19 \\
27 & 0.92 $\pm$ 0.21 & 1.09 $\pm$ 0.25 \\
29 & 0.63 $\pm$ 0.10 & 1.59 $\pm$ 0.25 \\
40 & 0.72 $\pm$ 0.12 & 1.39 $\pm$ 0.23 \\
46 & 0.70 $\pm$ 0.24 & 1.43 $\pm$ 0.49 \\
59 & 0.61 $\pm$ 0.16 & 1.64 $\pm$ 0.43 \\
110 & 0.77 $\pm$ 0.155 & 1.30 $\pm$ 0.26 \\
124	& 0.96 $\pm$	0.55 & 1.04 $\pm$ 0.60 \\
128 & 0.90 $\pm$	0.15 & 1.11 $\pm$ 0.19 \\
163	& 0.52 $\pm$	0.11 & 1.92 $\pm$ 0.40 \\
170 & 0.84 $\pm$	0.20 & 1.19 $\pm$ 0.28  \\
\hline
    \end{tabular}
    \label{tab:plxgaia}
\end{table}

\section{Analysis}  \label{4}
  
  In order to characterize the nature of the CRSs, we explored basic physical properties, namely spectral index, variability and polarization. The spectral index is a valuable preliminary diagnostic to assess whether the radio emission from a given source is of thermal or non thermal nature. Spectral indices below --0.1 indicate undoubtedly non-thermal nature \citep{rodriguez1993}. On the other hand, spectral indices above this limit are expected for thermal sources with the caveat that gyrosynchrotron emission can result in spectral indices up to 2.5. This latter emission implies non-thermal processes occurring usually on the stellar surfaces of young objects (e.g., magnetic reconnection events, \citealt{dulk1985,feigelson1999}). Besides, extragalactic sources are usually optically-thin emitters and their spectral indices can be mimicked by gyrosynchrotron sources \citep[e.g., ][]{fleishman2003,forbrich2016,wang2018}. However, as seen in the previous section, the number of extragalactic sources in our catalog of CRS is not statistically significant. As we discuss in Sect. 5, two important properties are associated with gyrosynchrotron emission: rapid variability and circular polarization.      
  
    \subsection{Spectral index}
  
  We estimated spectral indices of our CRSs dividing our data into two frequency intervals of 2 GHz and imaged each one in independent maps of 16 spectral windows. The maps have central frequencies of 9~GHz and 11~GHz, and noise levels of $11\,\mu$Jy~beam$^{-1}$ and $9\,\mu$Jy~beam$^{-1}$, respectively. Using BLOBCAT again, we determined the flux density of the CRSs belonging to each map and calculated their spectral indices. Most of the spectral indices were not well constrained due to the narrow frequency separation, causing large errors. Hence, in Table \ref{tab:spw} we list the sources whose spectral indices have errors $\leq 0.5$, as well as their flux density at 9 and 11 GHz. We note that negative spectral indices are dominant. Given that most sources are found at the center of the map (see Fig.~\ref{fig:bigim}), this can not be an effect of the primary beam difference between the 9 and 11 GHz maps. 
  
\begin{table}  
\caption{Spectral indices of the CRSs$^a$} \vspace{2mm}
\begin{center}
\setlength\tabcolsep{4pt}
\begin{tabular}{c c c c}
\hline \hline
        ID & $S_{9GHz}$ & $S_{11GHz}$ & Spectral \\
        \# & $(\mu Jy)$ & $(\mu Jy)$ & Index \\
            \hline
22 &  92486 $\pm$ 4626 & 111808 $\pm$ 5591 &  0.9 $\pm$  0.3\\
30 &  867 $\pm$ 48 & 573 $\pm$ 31 & --2.1 $\pm$  0.4\\
31 &  405 $\pm$ 29 & 499 $\pm$ 27 &  1.0 $\pm$  0.5\\
32 & 949 $\pm$ 55 & 791 $\pm$ 44 & --0.9 $\pm$  0.4\\
38 & 291 $\pm$ 21 &  222 $\pm$ 14 & --1.4 $\pm$  0.5\\
43 & 1526 $\pm$ 80 & 1814 $\pm$ 93 &  0.9 $\pm$  0.4\\
47 & 254 $\pm$ 18 & { 183} $\pm$ 11 & --1.6 $\pm$  0.5\\
69 & 656 $\pm$ 34 &  648 $\pm$ 33 & --0.1 $\pm$  0.4\\
79 & 885 $\pm$ 45 & 841 $\pm$ 43 & --0.3 $\pm$  0.4\\
84 & 250 $\pm$ 17 & 191 $\pm$ 12 & --1.3 $\pm$  0.5\\
86 & 519 $\pm$ 30 & 623 $\pm$ 34 &  0.9 $\pm$  0.4\\
94 & 632 $\pm$ 33 & 581 $\pm$ 30 & --0.4 $\pm$  0.3\\
99 &  439 $\pm$ 24 & 395 $\pm$ 21 & --0.5 $\pm$  0.4\\
105 &  241 $\pm$ 16 &  186 $\pm$ 11 & --1.3 $\pm$  0.4\\
108 &  304 $\pm$ 19 &  208 $\pm$ 12 & --1.9 $\pm$  0.4\\
110 &  194 $\pm$ 14 &  138 $\pm$ 9 &  --1.7 $\pm$  0.5\\
115 &  691 $\pm$ 36 &  701 $\pm$ 36 &  0.1 $\pm$  0.4\\
120 &  785 $\pm$ 41 & 770 $\pm$ 39 &  --0.1 $\pm$  0.4\\
151 & 692 $\pm$ 38 & 610 $\pm$ 33 & --0.6 $\pm$  0.4\\
152 &  893 $\pm$ 46 & 564 $\pm$ 29 & --2.3 $\pm$  0.4\\
159 &  369 $\pm$ 21 &  317 $\pm$ 17 & --0.8 $\pm$  0.4\\
161 &  164 $\pm$ 12 &  160 $\pm$ 10 & --0.1 $\pm$  0.5\\
\hline
\end{tabular}
\end{center}
\label{tab:spw}
$^a$We only report sources with spectral index errors smaller than or equal  to 0.5. \\
\end{table}

  \subsection{Rapid time variability}

We constructed maps for the data corresponding to different observing days (May 3, May 4 and May 8) to search for rapid variability in the CRSs. We applied BLOBCAT to each map using a signal to noise ratio of $3~\sigma$ because the individual maps are noisier than the map produced from the complete $uv$-data set. 

\begin{figure*}[!ht]
 \centering
  \subfigure[May $3$]{
    \includegraphics[height=0.22\textheight]{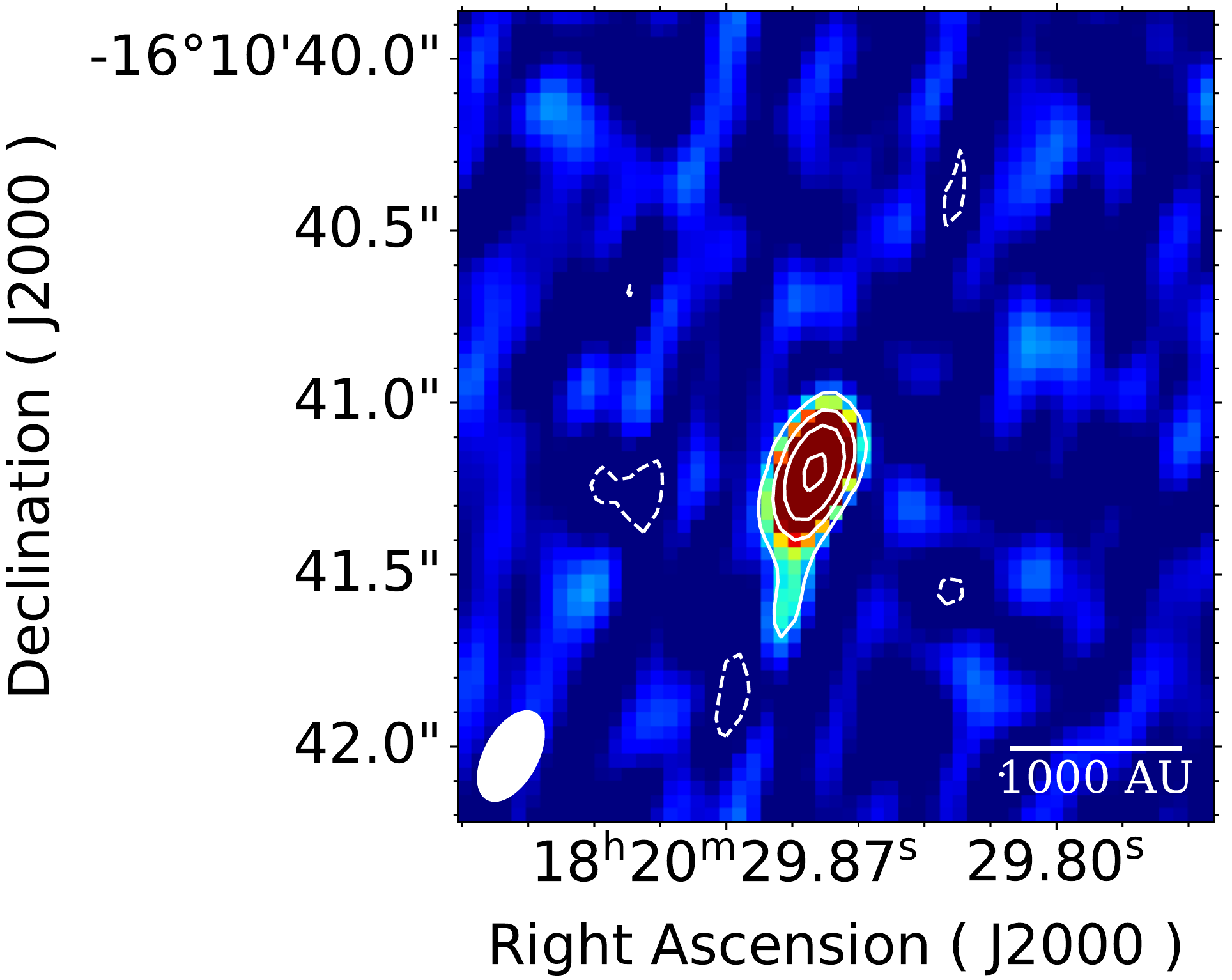}}
  \subfigure[May $4$]{
    \includegraphics[height=0.22\textheight]{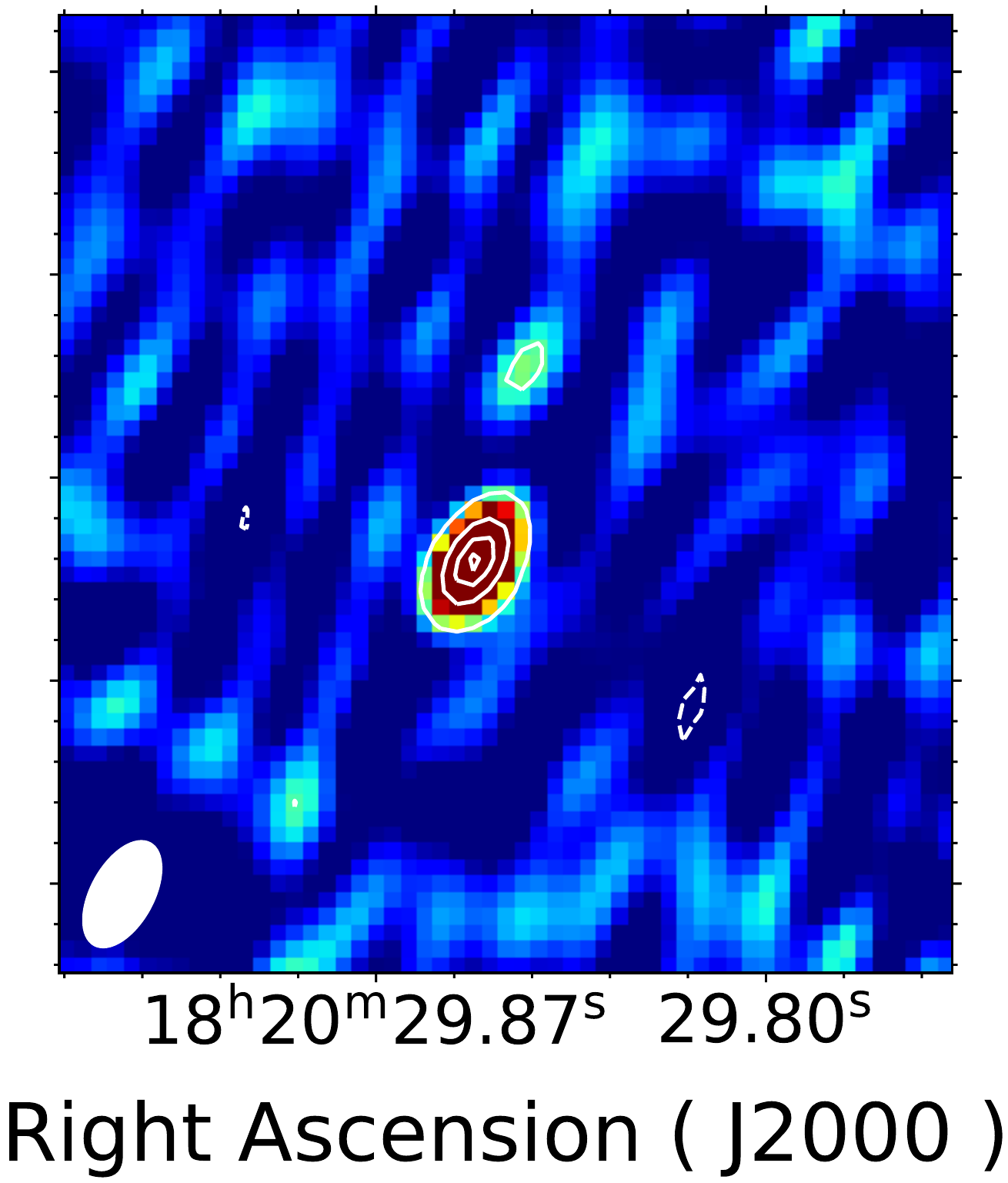}}
   \subfigure[May $8$]{
    \includegraphics[height=0.22\textheight]{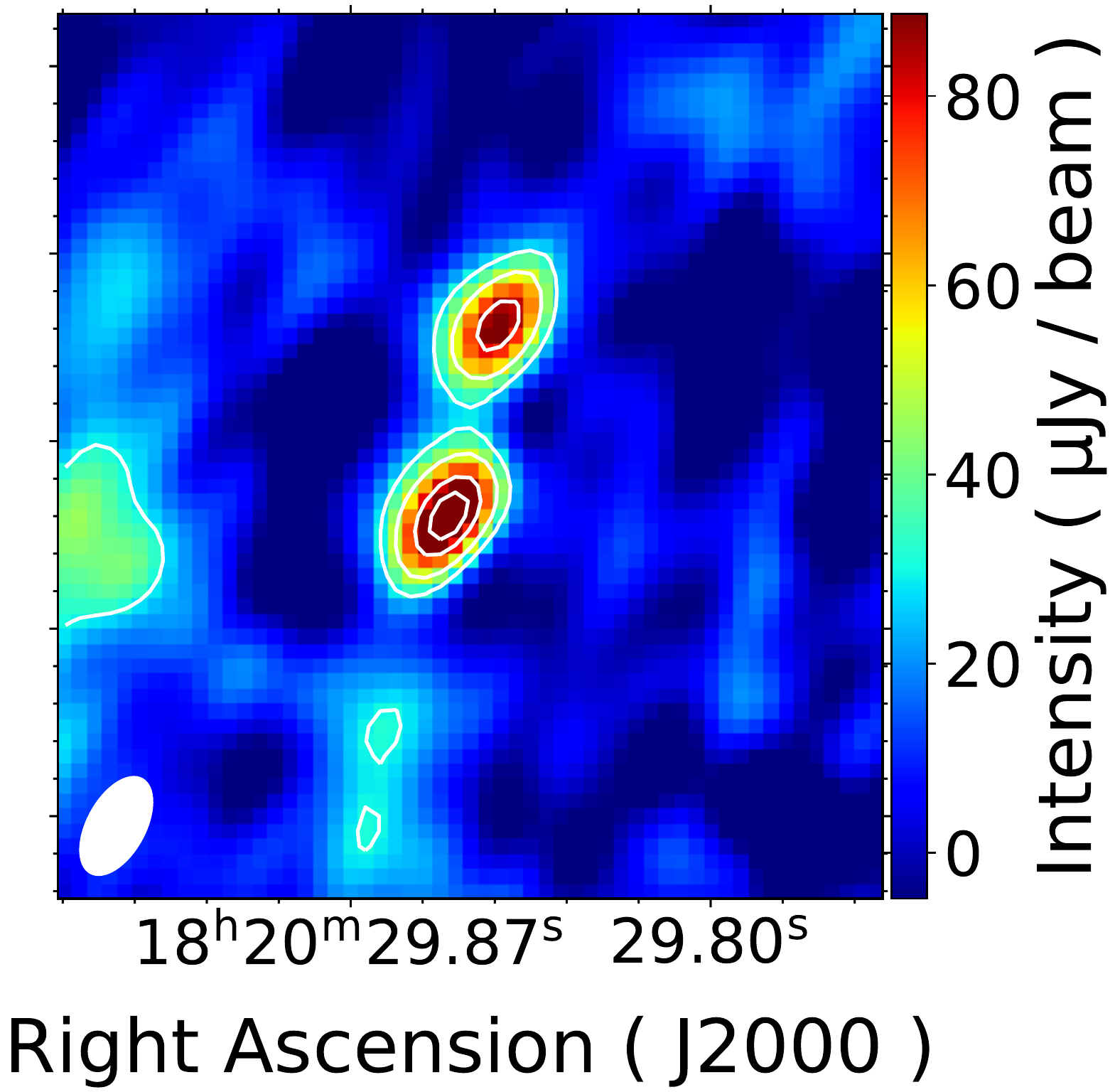}}
 \caption{Detailed view of rapid variability on two sources close to each other. The northern source is the number 82 and the southern source is the number 84. The color scale is the same in the three panels. In (a) we show the map on May 3. Contours are $-3, 3, 10, 20$ and $40$ times $8~\mu$Jy~beam$^{-1}$. In (b) we show the map on May 4. Contours are $-3, 3, 10, 15$ and $18$ times $12~\mu$Jy~beam$^{-1}$. In (c) we show the map on May 8. Contours are $-3, 3, 10, 15$ and $18$ times $12~\mu$Jy~beam$^{-1}$.}

 \label{fig:var}
\end{figure*}

The remaining faintest CRSs that BLOBCAT did not detect on single-day maps were measured using the task IMFIT. Then, we compared the fluxes of the CRSs between all the possible combinations of pairs of days by using the relation $V= (|S_{\rm day1} - S_{\rm day2}|\times100)/S_{\rm max}$, where $S_{\rm day1}$ and $S_{\rm day2}$ are the integrated fluxes on different days, and $S_{\rm max}$ is the larger of the two flux densities. We considered \textit{highly variable sources} those with $V\geq50\%$ \citep[e.g.,][]{dzib2013c}, considering that the average variability of sources that are not highly variable
is $30\%$. Table \ref{tab:highvar} shows a total of $72$ CRSs with rapid high variability, around $40\%$ of our catalog. Note that this fraction is a lower limit because some sources determined as non variable may be highly variable but be caught at an epoch in which their emission is steady. Assuming that our CRSs are YSOs, this lower limit is consistent with the result found in other regions \citep{dzib2013c,dzib2015,ortizleon2015}. The error in V was determined following standard error propagation theory. In Figure \ref{fig:var} we show an example of two CRSs close to each other that present rapid variability. The southern source decreases in luminosity in the lapse of a few days while the northern source exhibits the opposite behavior.
  
\begin{table*}[]
    \centering
    
    \caption{Rapid highly variable sources}
    \scalebox{0.92}{\begin{tabular}{c c c c | c c c c}
    \hline \hline
    ID & $May~3-4$ & $May~4-8$ & $May~3-8$ & ID & $May~3-4$ & $May~4-8$ & $May~3-8$\\
    \# & \% & \% & \% & \# & \% & \% & \% \\
    \hline 
2 &  19 $\pm$   4 &  55 $\pm$  12 &  44 $\pm$ 9 & 83 &  13 $\pm$   2 &  64 $\pm$   9 &  58 $\pm$   6 \\ 
3 &  40 $\pm$  10 &  75 $\pm$  15 &  59 $\pm$  14 & 84 &  41 $\pm$   4 &  42 $\pm$   5 &  66 $\pm$   7 \\
4 &  26 $\pm$   6 &  59 $\pm$  11 &  70 $\pm$  17 & 93 &   0 $\pm$   0 &  75 $\pm$  26 &  75 $\pm$  16 \\
6 &  63 $\pm$  18 &  54 $\pm$  17 &  19 $\pm$   5 & 96 &  $\geq$ 14 $\pm$   6 & $\geq$ 50 $\pm$  18 &  57 $\pm$  15 \\
9 &  $\geq$ 75 $\pm$  26 & $\geq$ 88 $\pm$  31 &  52 $\pm$   8 & 97 & $\geq$ 54 $\pm$  24 &  32 $\pm$  10 & $\geq$ 68 $\pm$  25 \\
11 &  54 $\pm$   9 &  11 $\pm$   3 &  49 $\pm$  11 & 101 &  35 $\pm$   9 & $\geq$ 64 $\pm$  25 & $\geq$ 44 $\pm$  17 \\
12 &  21 $\pm$   4 &  41 $\pm$   9 &  53 $\pm$  10 &  104 &  $\geq$ 24 $\pm$  10 &  $\geq$ 38 $\pm$  14 &  53 $\pm$  16 \\
14 &  17 $\pm$   2 &  60 $\pm$  11 &  67 $\pm$  11 & 106 & 64 $\pm$  23 & $\geq$ 61 $\pm$  24 & $\geq$ 10 $\pm$   4 \\
15 &  38 $\pm$  10 & $\geq$ 62 $\pm$  25 & $\geq$ 77 $\pm$  27 & 111 & $\geq$  59 $\pm$  26 &  34 $\pm$  11 & $\geq$ 73 $\pm$  26 \\
16 & $\geq$ 90 $\pm$  32 &  27 $\pm$   4 &  $\geq$ 93 $\pm$  32 & 114 & $\geq$ 17 $\pm$   7 & $\geq$ 55 $\pm$  19 & 62 $\pm$ 16 \\
19 &  52 $\pm$  14 &  45 $\pm$   8 &  14 $\pm$   3 & 117 & $\geq$ 43 $\pm$  20 & $\geq$ 75 $\pm$  26 & $\geq$ 86 $\pm$  29 \\
21 &   8 $\pm$   2 &  84 $\pm$  15 &  82 $\pm$  12 & 123 &   6 $\pm$   1 &  69 $\pm$  16 &  67 $\pm$  10 \\
23 &  30 $\pm$   5 & $\geq$ 77 $\pm$  28 &  $\geq$ 84 $\pm$  29 & 124 &   9 $\pm$   3 &  50 $\pm$  23 &  54 $\pm$  22 \\
24 &  44 $\pm$   4 &  66 $\pm$   6 &  39 $\pm$   3 & 125 &   7 $\pm$   2 &  66 $\pm$  17 &  63 $\pm$  11 \\
25 &  26 $\pm$   3 &  74 $\pm$  15 &  65 $\pm$  13 & 127 &   3 $\pm$   1 &  62 $\pm$  10 &  63 $\pm$   8 \\
26 &  23 $\pm$   8 & $\geq$ 41 $\pm$  18 & $\geq$ 55 $\pm$  21 & 131 &  88 $\pm$  15 &  65 $\pm$   7 &  66 $\pm$  12 \\
27 & $\geq$ 88 $\pm$  33 &  39 $\pm$  10 & $\geq$ 81 $\pm$  32 & 136 & $\geq$ 92 $\pm$  32 &  69 $\pm$  11 & $\geq$ 76 $\pm$  27 \\
32 &   7 $\pm$   1 &  50 $\pm$   4 &  54 $\pm$   5 & 139 &  35 $\pm$  17 &  57 $\pm$  23 &  72 $\pm$  25 \\
35 &   8 $\pm$   2 &  77 $\pm$  21 &  75 $\pm$  15 & 144 & $\geq$ 43 $\pm$  20 & $\geq$ 16 $\pm$   6 &  $\geq$ 52 $\pm$  20 \\
36 &  63 $\pm$  18 & $\geq$ 47 $\pm$  20 & $\geq$ 81 $\pm$  28 & 146 &  49 $\pm$  19 &  17 $\pm$   5 &  58 $\pm$  20 \\
40 & $\geq$  7 $\pm$   3 & $\geq$ 90 $\pm$  30 &  91 $\pm$  22 &  152 &   3 $\pm$   0.2 &  53 $\pm$   4 &  52 $\pm$   4 \\
42 &  22 $\pm$   6 & $\geq$ 61 $\pm$  24 & $\geq$ 50 $\pm$  18 & 158 &$\geq$  17 $\pm$   7 & $\geq$ 84 $\pm$  28 &  86 $\pm$  23 \\
46 &  37 $\pm$  13 & $\geq$ 32 $\pm$  15 & $\geq$ 57 $\pm$  20 & 162 &  17 $\pm$   7 &  68 $\pm$  25 &  61 $\pm$  13 \\
52 &  50 $\pm$  15 &  24 $\pm$   6 &  62 $\pm$  18 &  163 &  66 $\pm$  13 &  23 $\pm$   4 &  55 $\pm$  11 \\
54 &   7 $\pm$   2 & $\geq$ 52 $\pm$  22 & $\geq$ 55 $\pm$ 20 & 164 & 33 $\pm$  12 &  69 $\pm$  20 &  79 $\pm$  22 \\
56 &  34 $\pm$  16 &  25 $\pm$  11 &  51 $\pm$  18  & 167 &  30 $\pm$   9 &  76 $\pm$  19 &  66 $\pm$  17 \\
57 & $\geq$ 10 $\pm$   4 & $\geq$ 58 $\pm$  21 &  62 $\pm$  16 & 170 &  36 $\pm$   9 &  24 $\pm$   7 &  52 $\pm$  10 \\
61 &  27 $\pm$  13 &  48 $\pm$  20 &  62 $\pm$  18 & 171 & $\geq$ 60 $\pm$  40 & $\geq$ 77 $\pm$  26 &  91 $\pm$  54 \\
62 &  33 $\pm$  11 &  37 $\pm$  11 &  57 $\pm$  15 & 173 &   0 $\pm$   0 &  88 $\pm$  11 &  88 $\pm$  16 \\
65 &  15 $\pm$   5 & $\geq$ 55 $\pm$  23 & $\geq$ 46 $\pm$  17 & 174 &  17 $\pm$   3 &  52 $\pm$  10 &  42 $\pm$   5 \\
66 &  61 $\pm$  18 &  45 $\pm$   9 &  78 $\pm$  20 & 175 & $\geq$ 47 $\pm$  16 & $\geq$ 83 $\pm$  29 &  68 $\pm$  10 \\
71 &  52 $\pm$   8 &  85 $\pm$  10 &  93 $\pm$  12 & 177 &  $\geq$ 41 $\pm$  14 & $\geq$ 51 $\pm$  18 &  16 $\pm$   3 \\
73 &   5 $\pm$   2 &  65 $\pm$  18 &  63 $\pm$  11 & 178 & $\geq$ 87 $\pm$  31 &  70 $\pm$  15 & $\geq$ 56 $\pm$  21 \\
78 &  17 $\pm$   4 & $\geq$ 61 $\pm$  24 & $\geq$ 53 $\pm$  19 & 179 &  12 $\pm$   6 &  53 $\pm$  28 &  59 $\pm$  13 \\
82 & $\geq$ 51 $\pm$  22 &  56 $\pm$  17 & $\geq$ 78 $\pm$  28 & 182 &  53 $\pm$  23 &  16 $\pm$   7 &  44 $\pm$  10 \\
\hline
    \end{tabular} }
    \label{tab:highvar}
\end{table*}

  \subsection{Polarization}
  
  We searched for circular polarization, which is indicative of gyrosynchrotron emission. To this end we used the CASA task CLEAN to map the Stokes V parameter. The rms noise was $4~\mu$Jy~beam$^{-1}$ in the central region. Stokes V and I are measures of the circular polarization and total intensity (i.e. final map of Sect 2), respectively, of the radiation, where the V/I flux ration gives the degree of circular polarization. The polarization was considered significant only when the sources are close to the phase center ($<1'$; sources more distant from the phase center suffer from the 'beam squint' effect that produces spurious circular polarization) and the flux density ratio ($S_{\rm peakV}\times 100/S_{\rm peakI}$) is greater than $3\%$, where $S_{\rm peakV}$ is the peak flux density from the Stokes V map and $S_{\rm peakI}$ is the peak flux density obtained in the Stokes I map. Table \ref{tab:stokes} shows $S_{\rm peakV}$ and $S_{\rm peakI}$ of the CRSs that fulfill these criteria and their corresponding flux density ratio. The error in the flux density ratio was determined following standard error propagation theory.

\begin{table}[t]
    \centering
    \caption{Circularly polarized sources.}
    \begin{tabular}{c   c   c   c }
        \hline \hline
        ID & $S_{peakV}$ & $S_{peakI}$ & Flux density ratio  \\
        \# & ($\mu$Jy~bm$^{-1}$) & ($\mu$Jy~bm$^{-1}$) & \% \\
        \hline 
        56 & 17 $\pm$ 3 & 42 $\pm$ 6 & 40 $\pm$ 9 \\
        62 & --16 $\pm$ 4 & 34 $\pm$ 6 & 47 $\pm$ 5\\
        95 & 17 $\pm$ 4 & 79 $\pm$ 7 & 21 $\pm$ 5\\
        113 & --14 $\pm$ 5 & 46 $\pm$ 6 & 30 $\pm$ 12 \\
        122 & 16 $\pm$ 5 & 61 $\pm$ 6 & 26 $\pm$ 9\\
        \hline
    \end{tabular}
    
    \label{tab:stokes}
\end{table}

\section{Discussion} \label{5}

The extraordinary large number of radio sources found in M\,17 is possibly related with the hundreds of sources reported in previous surveys at other wavelengths \citep[e.g., 771 X-ray sources with stellar counterparts and 96 candidate YSOs:][]{broos2007,povich2009}, as suggested by the counterparts reported in Sect. 3.4. As pointed out by \citetalias{rodriguez2012}, their detected radio sources probably correspond to the same class of objects detected at other wavelengths and any lack of counterparts is caused by heavy extinction. In order to confirm the same trend in our CRSs, in Sect. 4 we explored some source properties, from which we attempt to unveil their emission mechanism. These properties will be discussed in this section and compared, when possible, to what is found for other regions. 

\begin{figure*}[h!]
    \centering
    \includegraphics[width=0.8\textwidth]{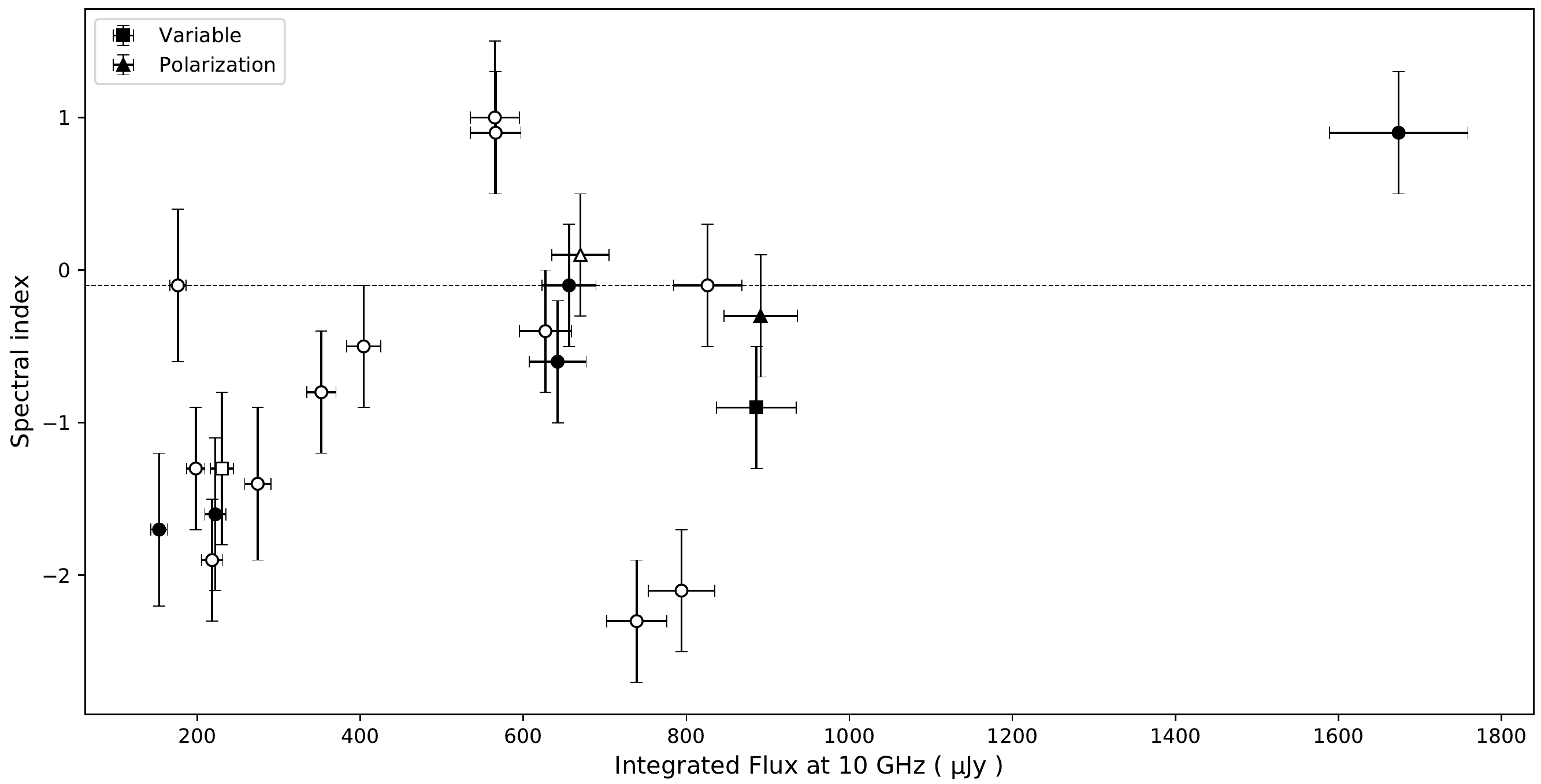}
    \caption{Spectral index against integrated flux density at 10 GHz of the brightest  CRSs. Squares are the highly variable sources. Triangles are the sources with polarization. The filled symbols are those with X-ray counterpart and the unfilled are the sources without an X-ray counterpart. The UC1 source is not included in the plot. The dashed line marks a spectral index of $-0.1$.}
    \label{fig:alphavsflux}
\end{figure*}

In Fig. \ref{fig:alphavsflux} we plot the spectral index vs. flux density for our brighter CRSs. Other than the predominance of negative spectral indices highlighted in the previous sections, there are no significant trends in the diagram. 

Given the deeper detection level achieved by \citet{forbrich2016} in Orion, a significant trend was revealed, where the brighter sources in Orion (at least a few mJy) have positive spectral indices, while the sources with negative spectral indices are significantly weaker. This behavior is a consequence of the smaller emitting region of non-thermal emission in YSOs, usually limited to the stellar corona. At the same time, this implies that non-thermal emission associated to stellar objects usually appears point-like.  
While the presence of thermal sources of a few mJy in Orion is consistent with the 1.7 mJy source in M\,17 seen in the diagram (scaled by distance), our detection limit prevents us from searching for the aforementioned trend for fainter sources. However, while the highly variable CRSs corresponds to only 10\% of the CRSs
plotted in Figure \ref{fig:alphavsflux}, Table 6 shows that the fraction of highly variable CRSs represents 40\% of our catalog, indicating that most of the highly variable CRSs are faint objects. Although variablity is likely associated to non-thermal emission, additional factors are ususally required. If we assume that the trend of \citet{forbrich2016} also applies to the sources of M\,17, then variability could be tied to non-thermal emission, at least for a subtantial part of our CRSs (most of which do not appear in the graph).

Gyrosynchrotron radiation from stellar coronae is a common type of point-like, non-thermal emission found in regions of young stellar objects. In the case of gyrosynchrotron emission associated with low-mass stars, electrons accelerated during magnetic reconnection events are responsible both for the radio emission when they propagate in the stellar magnetic field, and the X-ray emission through their heating of the coronal gas \citep{benz2010}. This type of emission can present variability and also shows a significant degree of cicular polarization. Therefore, from the total catalog of CRSs, we identify  polarized and variable sources as likely gyrosynchrotron candidates \citep[e.g.,][hereafter referred to as Gy-CRS]{dulk1985,Gudel2002,dzib2013c,dzib2015,ortizleon2015}. We consider polarization and variability as complementary since some CRSs might be variable but present flux density stability during the short periods covered by our observations. 

The rest of the CRSs cannot be uniquely categorized and we leave their nature as undetermined (hereafter referred to as Und-CRS). A small fraction of Und-CRS might be thermal emitters and they could be small trapped HCHII regions or proplyds. Moreover, some Und-CRS have IR counterparts that could indicate the presence of heated dust. 
The typical fluxes detected by us ($< 1$ mJy) are similar to those of proplyds detected in Orion \citep[][]{zapata2004} scaled to the M\,17 distance and, thus, we favor the proplyd scenario. Other Und-CRS could be non-thermal emitters that present low or no variability (e.g., collision wind region between massive stars where electrons are highly accelerated). Observations at higher frequencies with higher angular resolution are needed to explore the nature of Und-CRS.      

\subsection{The X-ray--Radio connection}

As we mentioned earlier, it is thought that the X-ray and radio emissions in stellar coronas are both caused by electrons accelerated during magnetic reconnection events. The electrons accelerated during these events generate the radio emission via the gyro-synchrotron mechanism, but they also heat the corona to temperatures sufficient to produce thermal bremsstrahlung X-ray emission \citep{dulk1985,feigelson1999,Gudel2002}. This correlation  between radio and X-ray emission was first investigated empirically by \citet{Gudel1993} and \citet{benz1994} for a population of coronally active stars and this relation was found:

\begin{equation}
\frac{L_X}{L_R}=\kappa 10^{15.5\pm0.5} 
\end{equation}

\noindent with $\kappa \leq 1$ accounting for different types of stars. 
For instance, several works show that $\kappa$ is significantly lower than 1 for YSOs \citep{gagne2004,dzib2013c,dzib2015}.

\begin{figure*}[!ht]
 \centering
 \includegraphics[height=0.5\textheight]{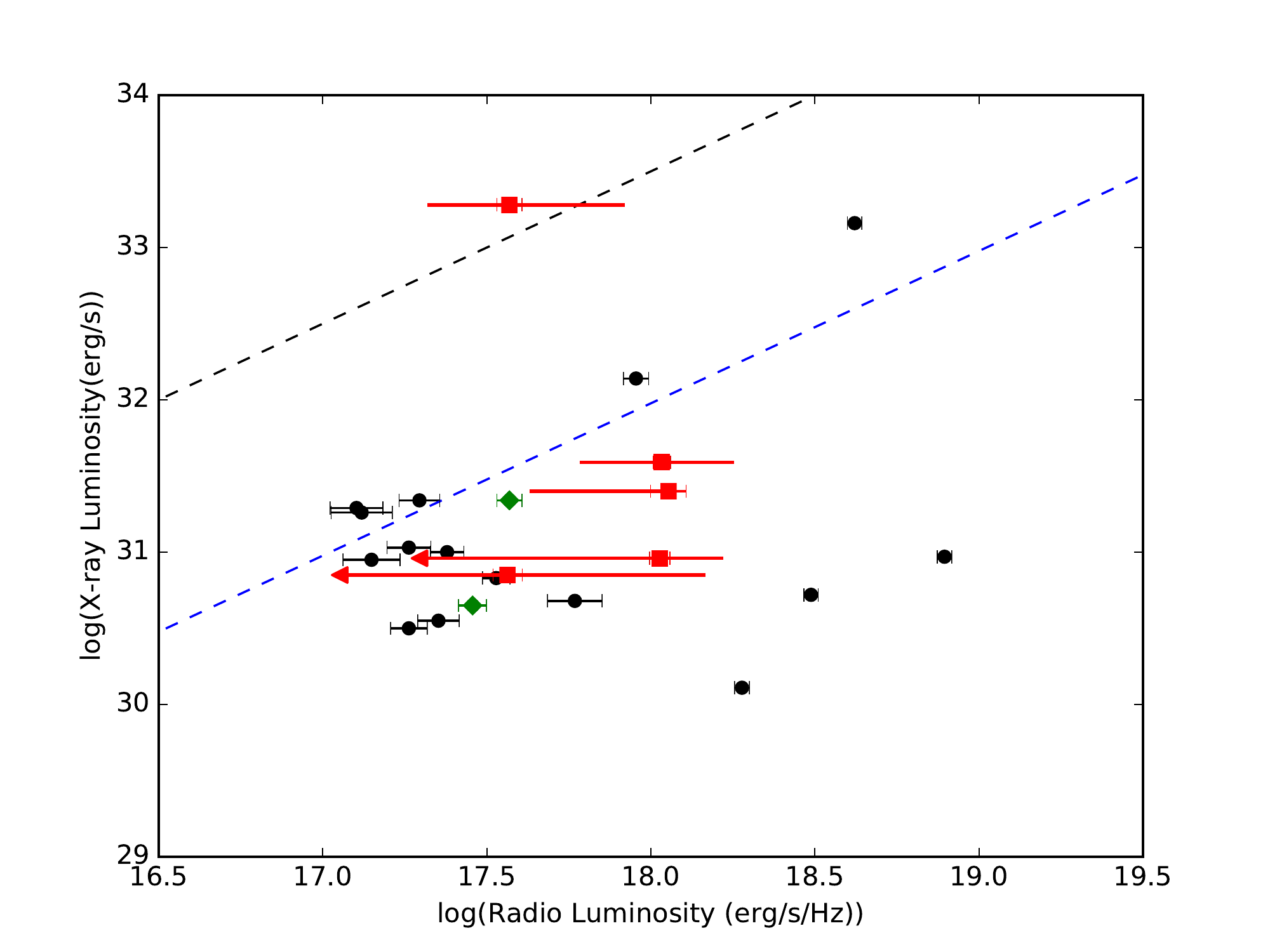}
 \caption{X-ray luminosity vs. Radio luminosity at 10 GHz of the CRSs in M\,17. Black dots represents the Und-CRS, green diamonds are sources with circular polarization and red squares are variable sources (these two types of sources are Gy-CRS), with the horizontal red bars representing the range of variability between our observing epochs. If the variable source is undetected in one or more epochs, we report upper limits as arrows at the lower end of the range. The black solid and blue dashed lines represent the G\"udel-Benz relation with $\kappa$ = 1 and 0.03, respectively, the latter employed by \citet{dzib2013c} to match radio and X-ray luminosites of YSOs.
 \label{fig:gb}}
\end{figure*}

From our catalog, 24 of 182 CRSs have X-ray counterparts with measured luminosities reported in \citet{broos2007}, and some of them are absorption corrected. They followed a similar methodology as that in \citet{feigelson2002}: wide band fluxes were obtained and converted to total-band luminosities (0.5-8 kev) by fitting a thermal plasma model to the data under the maximum likelihood method. The fit also gives the absorbing column of interstellar material and, thus, corrected luminosities by absorption can  also be obtained. When the fit yielded unrealistic physical parameters, Broos et al. employed a power-law model to fit the data. Besides, some corrected luminosities measured only the hard X ray band
(2-8 kev) and hence are meaningless because they correspond to highly obscured sources. Thus, the absorption-corrected luminosities are omitted for these sources and only their observed luminosity are reported. The X-ray luminosities and our derived radio luminosities for CRSs in M\,17 are shown in Figure \ref{fig:gb}. 
In this plot, we excluded those Gaia counterparts not belonging to the region (i.e., foreground sources). A linear least squares fit to the 21 remaining data points shown in Figure \ref{fig:gb} gives: 

\begin{center}

log $L_\mathrm{X}$ = ($26.6\pm5.7$) + ($0.26\pm0.32$) log $L_\mathrm{R}$

\end{center}

\noindent with a correlation coefficient (R) of 0.17. Since this coefficient has a low value, we conclude that the radio and X-ray luminosities are not correlated and, thus, do not follow a G\"udel-Benz relation, where a slope consistent with 1 is expected. Note that, in general, our data are underluminous in X-ray emission. Lower X-ray luminosities than those predicted by the G\"udel-Benz relation are also found in other star forming regions regions \citep{kounkel2014, dzib2013c, ortizleon2015, dzib2015,forbrich2016}, meaning that we are likely picking up objects of the same nature. Also, Gy-CRS and Und-CRS follow the same behavior and there is no segregation between them.  

The lack of correlation between X-ray and radio data can be due to several reasons. First, our data sample is contaminated by a number of possible thermal sources because points from the undetermined category (black dots) are dominant in the graph, while the G\"udel-Benz relation only applies for non-thermal sources. Second, our sample is small and CRSs with luminosities below log ($L_\mathrm{R}$) $\sim17.0$ erg s$^{-1}$Hz$^{-1}$ are not detected, perhaps due to a lack of completeness, adding uncertainty in the slope determination. This is not surprising since M\,17 is further away than other nearby regions observed in similar studies \citep{dzib2013c, ortizleon2015} and suggests that the faintest objects of our region remain undetected by our observations.    

\begin{figure*}[h]
 \centering
 \includegraphics[height=0.6\textheight]{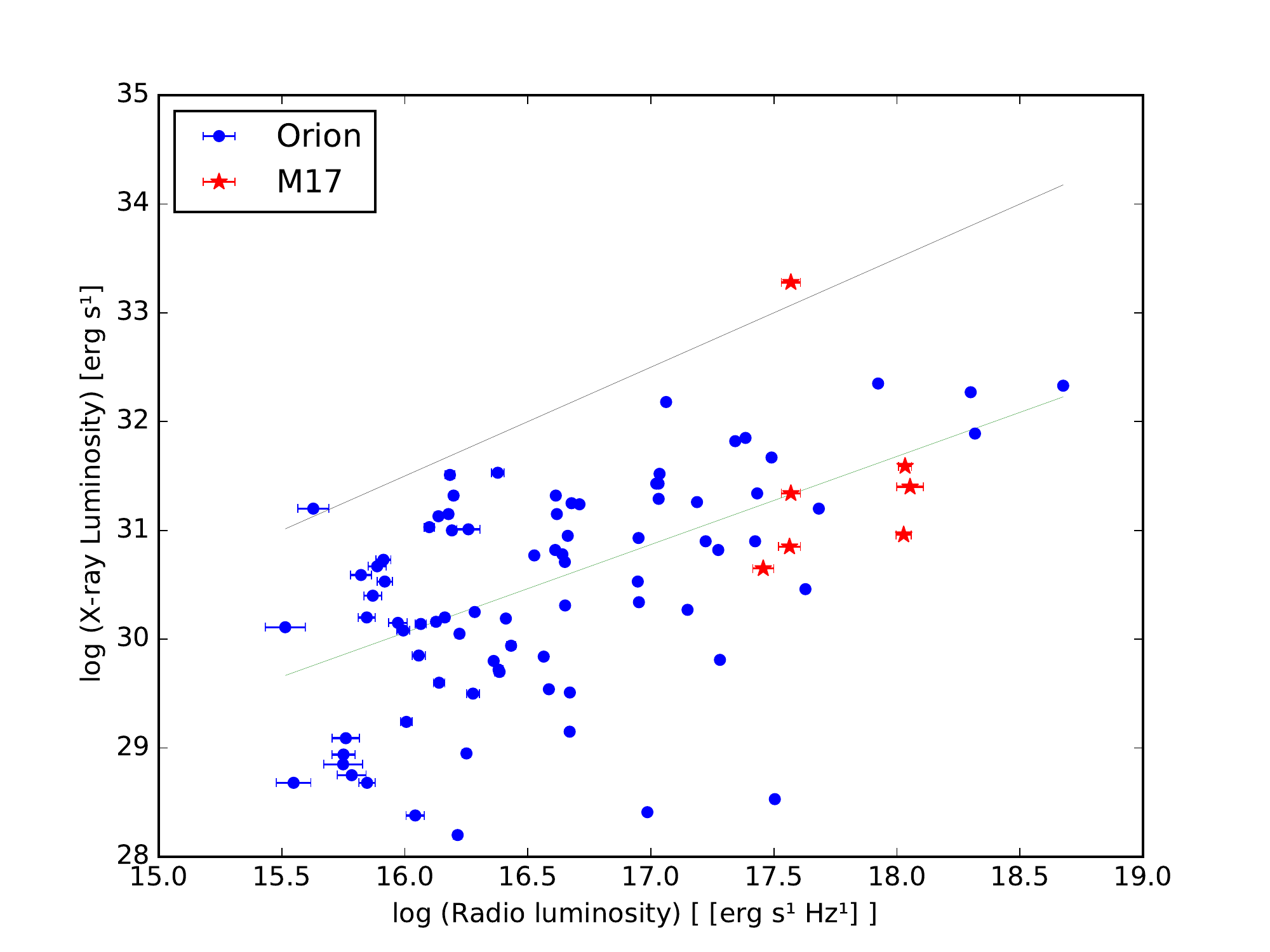}
 \caption{X-ray luminosities as a function of radio luminosity. The points mark the position of the 82 X-ray sources that are also detected with VLBI techniques (blue dots). The red stars are data points of the M\,17 region with X-ray emission and non-thermal characteristics in the radio. The green line is the fit discussed in the text. The black line is the original G\"udel-Benz relation.}\label{fig:GB_OriM17}
\end{figure*}

As discussed by Forbrich et al. (2016), a similar trend is observed for the 250 ONC sources with both X-ray and radio emission. If we fit their respective luminosities we obtain:

\begin{center}

log $L_\mathrm{X}$ = ($24.1\pm1.6$) + ($0.37\pm0.10$) log $L_\mathrm{R}$

\end{center}

\noindent again suggesting that the X-ray and radio luminosities are poorly correlated, if at all (R$ =0.23$). This is not unexpected since a large fraction of the radio sources in the ONC are thermal and their intensity is unrelated to the mildly relativistic electrons that produce the X-ray and gyrosynchrotron emission in active stars (G\"udel  \& Benz 1993). If, in contrast, for Orion we consider only the 82 sources with X-ray emission that are also VLBI sources \citep{forbrich2021,dzib2021} and whose emission is thus clearly non-thermal\footnote{We consider a VLBI source as a non-thermal emitter since the VLBI technique filters out the thermal emission leaving only the most compact emission with high brightness temperatures ($>10^6$ K), a strong indicator of non-thermal emission associated to young stars \citep[e.g.,][]{feigelson1999}.}, a clear correlation is obtained. These data are fitted by:

log $L_\mathrm{X}$ = ($17.1\pm2.4$) + ($0.81\pm0.14$) log $L_\mathrm{R}$.

\noindent consistent with a linear fit (R$ =0.53$). The data and fit are shown in Figure~\ref{fig:GB_OriM17}. The obtained fit for this case is also about an order of magnitude lower in $L_\mathrm{X}$ than the original G\"udel-Benz relation. This result confirms the notion that the radio emission is probably produced by gyrosynchrotron radiation from the same mildly relativistic electrons that are responsible for the X-ray emission. Unfortunately, the same test for the M\,17 data yields unreliable results because we have only 8 sources with X-ray emission and non-thermal characteristics in the radio. Nevertheless, we have plotted this handful of sources in Figure \ref{fig:GB_OriM17} to show that they fall near the high-brightness portion of the Orion data. We speculate that more sensitive observations of M\,17 will reveal a similar population of related radio - X-ray sources. As discussed below, for a given radio luminosity M\,17 has about three times the number of sources of Orion. In practice, however, given its larger distance it will be very difficult to obtain an M\,17 survey with similar luminosity limits as those obtained for Orion.

\subsection{The luminosity function} 

\begin{figure}[H]
 \centering
 \includegraphics[height=0.4\textwidth]{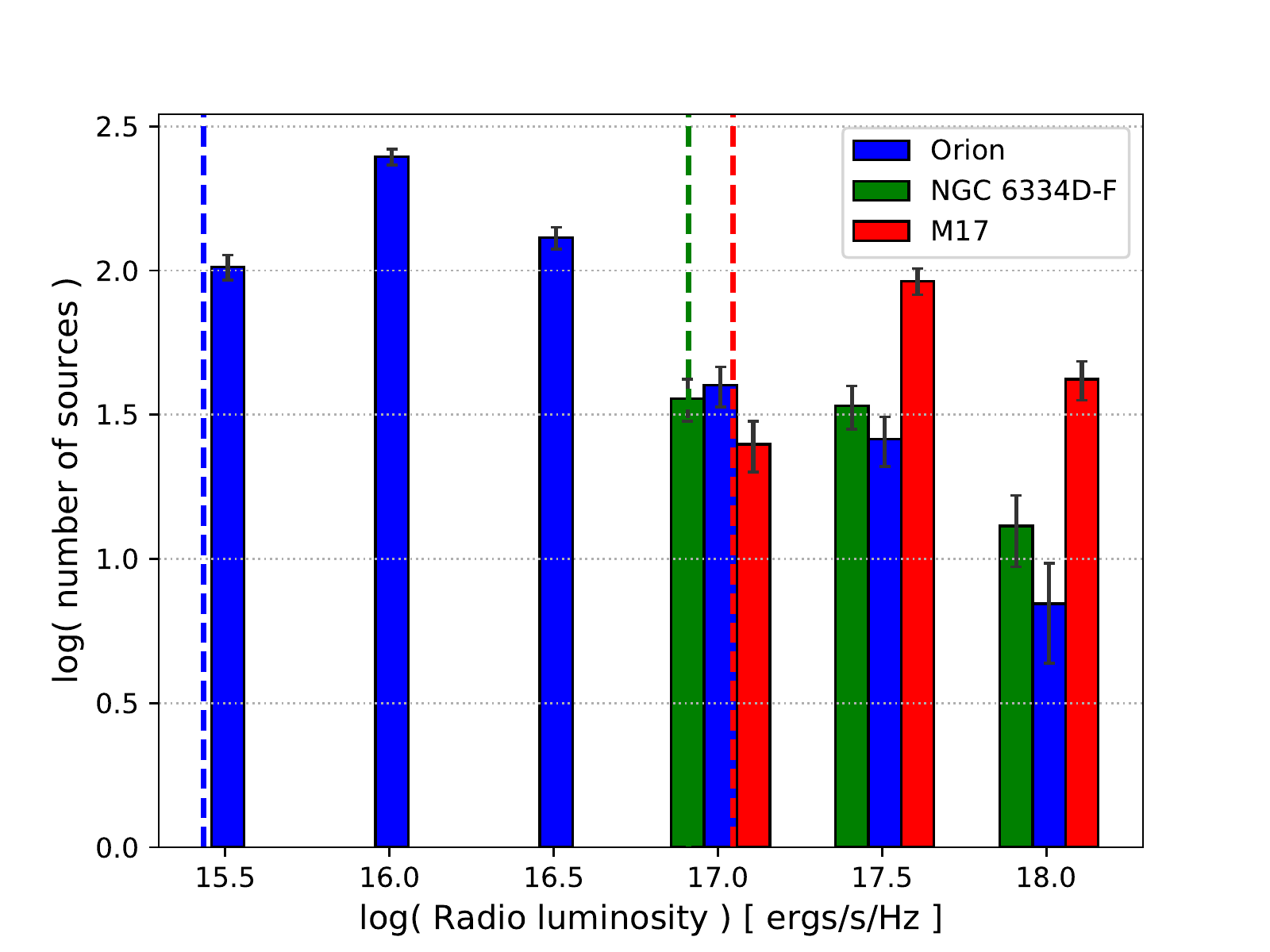}
 \caption{Radio luminosity distribution of M\,17 (red bars, this work),  NGC\,6334D-F (green bars, \citealt{medina2018}) and Orion (blue bars, \citealt{forbrich2016}). The error bars are the uncertainty assuming that the number of sources in each luminosity bin follows a Poisson distribution. The vertical dashed lines indicate the adopted detection limit for each region according to the respective color. Note that the position of these lines makes the log(Radio luminosity) = 17.0 erg/s/Hz bins incomplete for M\,17 and NGC\,6334D-F. The width of each bin is 0.5 dex.
 \label{fig:comp}}
\end{figure}

The high number of detected CRSs towards M\,17 (182 CRSs) makes this region the second richest in radio source detections after the ONC (556 CRSs), and followed by NGC\,6334 (83 CRSs). Thus, a comparison between these regions is tempting. 
In Figure \ref{fig:comp} we show histograms for the radio luminosity distribution (number of CRSs per logarithmic luminosity interval) of the three regions of star formation considered here. Unfortunately, because of their larger distances, we are severely limited in the dynamic range for M\,17 and NGC\,6334D-F due to  incompleteness, and a comparison can be made only for the two bins of higher luminosity. 

M\,17 has about three times more sources
at a given radio luminosity than the other two regions. This result is probably related with the significantly higher global luminosity (stars plus dust) that M\,17 has over the two
other regions (Johnson 1973). We also conclude that the negative slope between these two luminosity bins is, within error, the same for all three regions. We find that, approximately, the number of sources drops inversely (in a linear relation) with their radio luminosity.

\subsection{Age and spatial distribution of the CRSs}

The VLA observations from the Gould's Belt Very Large Array Survey
studied the radio emission from hundred of YSOs in nearby ($<500$~pc) star-forming regions \citep{dzib2013a,dzib2015,kounkel2014,ortiz-leon2015,pech2016}.
The results have shown that the properties of the radio emission (variability, spectral index, and circular polarization) change with the infrared classification of YSOs. Class 0,
and Class I young stars are mostly associated with thermal free-free radio emission, 
Class III with gyrosynchrotron radio emission and Class II is an intermediate case. However, some Class I YSOs have been confirmed to be gyrosynchrotron radio emmiters \citep[e.g.,][]{dzib2010}. 
Thus, gyrosynchrotron sources usually correspond to relatively evolved young stars with clean surroundings and
their photosphere visible at optical wavelengths (Class III YSOs). 
On the other hand, thermal sources are likely younger objects still embedded in large envelopes of circumstellar material that are strongly accreting and at the same time producing jets whose free-free emission is dominant. The large number of gy-CRSs detected by us (36\% of the total CRSs, twice the VLBI sources found in Orion: $\sim20$\% of the total radio sources) suggests the existence of an important evolved population of young objects in the M\,17 region. The Und-CRS, on the other hand, probably include young thermal sources.

In the past, objects of different ages have been found in M\,17. For instance, \citet{nielbock2001} analyzed 22 compact NIR sources and deduced that most of them are massive
stars of Class I type, indicating that these objects formed very recently. In contrast, \citet{broos2007} analyzed 138 stars and found that only 12\% are Class
I and II, indicating a predominance of older Class II and III sources toward M\,17. 
Our results agree with the scenario of an age dispersion for the objects in M\,17 and suggest that the region has been forming stars at least ten million years (i.e. the age of Class III objects).

\begin{figure*}[h]
 \centering
 \includegraphics[height=0.6\textheight]{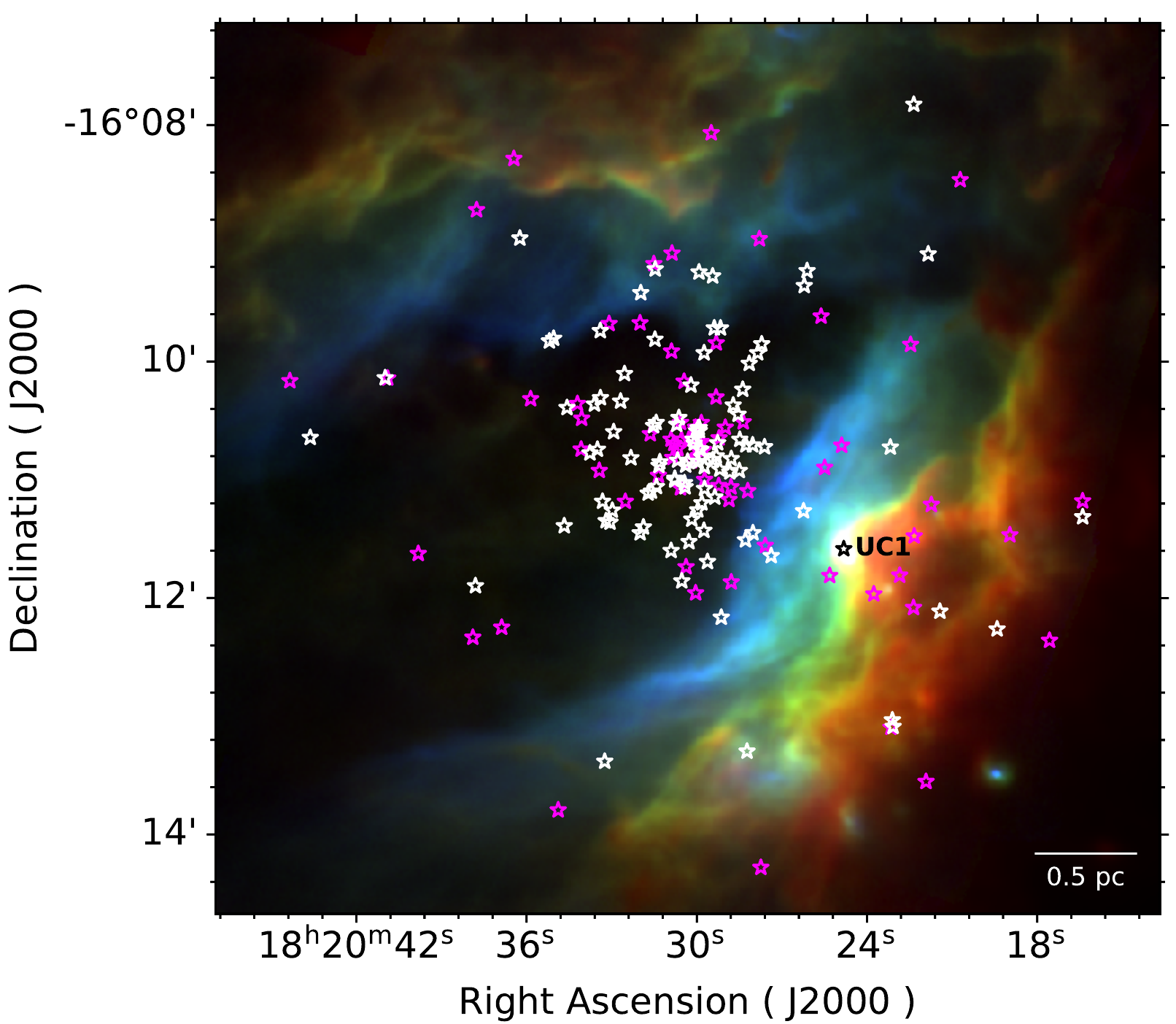}
 \caption{RGB image toward M\,17 of $\sim 7\arcmin \times 7\arcmin$ of field. Blue corresponds to the SOFIA-FORCAST $20~\mu$m image, green is the SOFIA $37~\mu$m image and red is the Herschel $70~\mu$m image. Pink stars represents the position of the Gy-CRS and white stars are the Und-CRS. The black star shows the UC1 position. The SOFIA images are taken from \citet{lim2020}.
 \label{fig:spatial_distribution}}
\end{figure*}

In Figure \ref{fig:spatial_distribution} we show the distribution of CRSs in the M\,17 region highlighting the Gy-CRSs in pink. The distribution of the two populations looks centrally peaked and smoothly fades in the outskirts of the map. Thus, to analyze the spatial distribution of radio sources in M\,17 we have fitted the data to a Gaussian ellipsoid following \citet{rodriguezgomez2019}. In Table \ref{tab:average_sigma} we present the parameters (centroid, major and minor axis and position angle of the major axis) for Gaussian ellipsoids fitted to the three populations of radio sources in M\,17 and in Orion, aiming to compare any segregation between populations in both regions. Note that the axes of the distributions are significantly smaller than the primary beam size. 

We conclude that for the case of Orion all three populations have parameters that overlap to within one-sigma. Perhaps this result is related to the lack of expansion or contraction motions found by \citet{dzib2017} from observations of 88 young stars with compact radio emission. It also suggests that, despite their age, the stars will remain similarly distributed.

For M\,17, the centroids are consistent within the noise and fall within a few arcsec of the positions of the massive binary system formed by CEN 1a and CEN 1b. Also the position angles are consistent within one-sigma. However, the major and minor axes of the non-thermal population seem to be about 20 percent larger than those of the other two populations, and significantly smaller than the primary beam size. The rough coincidence of the centroids of all three distributions in M\,17 with CEN 1 is consistent with a scenario where the potential well of the cloud centers on this massive binary star, and most of the objects (including the most massive) are formed around this position. Besides, the larger dispersion of the Gy-CRS is marginal and a direct measurement of their proper motions is required to provide a conclusive statement.     

We finally note that the radio population in M\,17 is not only about three times more abundant (a result inferred from the comparison of the most luminous sources) than in Orion, but also that it is more extended spatially. Correcting for the distance \citep[$388\pm5$~pc,][]{kounkel2017}, the major axis of the distribution of all sources in Orion corresponds to 0.17 pc, while it is 0.84 pc for the case of M\,17.

\begin{table*}[h]
\small
\centering
	\caption{Statistical parameters of the spatial distribution of CRSs}
\setlength\tabcolsep{4pt}
\begin{tabular}{l c c c c c}
\hline \hline
        CRS Type &   &   &     \multicolumn{3}{c}{Size} \\
         &    \multicolumn{2}{c}{Mean Position}    & Major Axis & Minor Axis & Position Angle  \\
         &  $\alpha$~(J2000) &   $\delta$~(J2000) & ($''$) & ($''$) & (deg.)  \\      
            \hline
\multicolumn{6}{|c|}{M17} \\
\hline
All CRS     &  $18^\mathrm{h}20^\mathrm{m}29\rlap{.}^\mathrm{s}84 \pm 0\rlap{.}^\mathrm{s}33 $ & $-16^{\circ}10'47\rlap{.}''9 \pm 4\rlap{.}''5 $& 66.9 $\pm$ 3.8  & 55.2 $\pm$ 3.0 & 52.0 $\pm$ 12.6   \\
Gy-CRS   &  $18^\mathrm{h}20^\mathrm{m}29\rlap{.}^\mathrm{s}54 \pm 0\rlap{.}^\mathrm{s}61 $ & $-16^{\circ}10'49\rlap{.}''7 \pm 8\rlap{.}''2 $& 79.2 $\pm$ 6.7  & 61.9 $\pm$ 5.2 & 53.6 $\pm$ 19.1   \\
Und-CRS &  $18^\mathrm{h}20^\mathrm{m}30\rlap{.}^\mathrm{s}04 \pm 0\rlap{.}^\mathrm{s}37 $& $-16^{\circ}10'46\rlap{.}''6  \pm 5\rlap{.}''3 $ & 56.6 $\pm$ 3.9  & 50.0 $\pm$ 3.5 & 47.8 $\pm$ 16.3  \\
\hline
\multicolumn{6}{|c|}{Orion Nebula Cluster} \\ 
\hline 
All sources     &  $5^\mathrm{h}35^\mathrm{m}16\rlap{.}^\mathrm{s}36 \pm 0\rlap{.}^\mathrm{s}21 $ & $-5^{\circ}22'55\rlap{.}''3 \pm 3\rlap{.}''4 $& 87.1 $\pm$ 2.6  & 67.7 $\pm$ 2.0 & -17.4 $\pm$ 7.1   \\
VLBI radio sources$^a$  &  $5^\mathrm{h}35^\mathrm{m}16\rlap{.}^\mathrm{s}45 \pm 0\rlap{.}^\mathrm{s}45 $ & $-5^{\circ}23'2\rlap{.}''2 \pm 7\rlap{.}''8 $& 90.0 $\pm$ 5.8  & 69.0 $\pm$ 4.4 & -28.7 $\pm$ 15.1  \\
non-VLBI radio sources$^b$ &  $5^\mathrm{h}35^\mathrm{m}16\rlap{.}^\mathrm{s}34 \pm 0\rlap{.}^\mathrm{s}22 $& $-5^{\circ}22'53\rlap{.}''4  \pm 4\rlap{.}''1 $ & 86.4 $\pm$ 2.9  & 66.9 $\pm$ 2.3 & -14.0 $\pm$ 8.0  \\
\hline
    \label{tab:average_sigma}
\end{tabular}
$^a$\citet{forbrich2021} and \citet{dzib2021} \\
$^b$\citet{forbrich2016} \\
\end{table*}

\subsection{The Hyper-Compact HII region UC1 and the Arc structure}

\begin{figure*}[h]
 \centering
 \includegraphics[height=0.4\textheight]{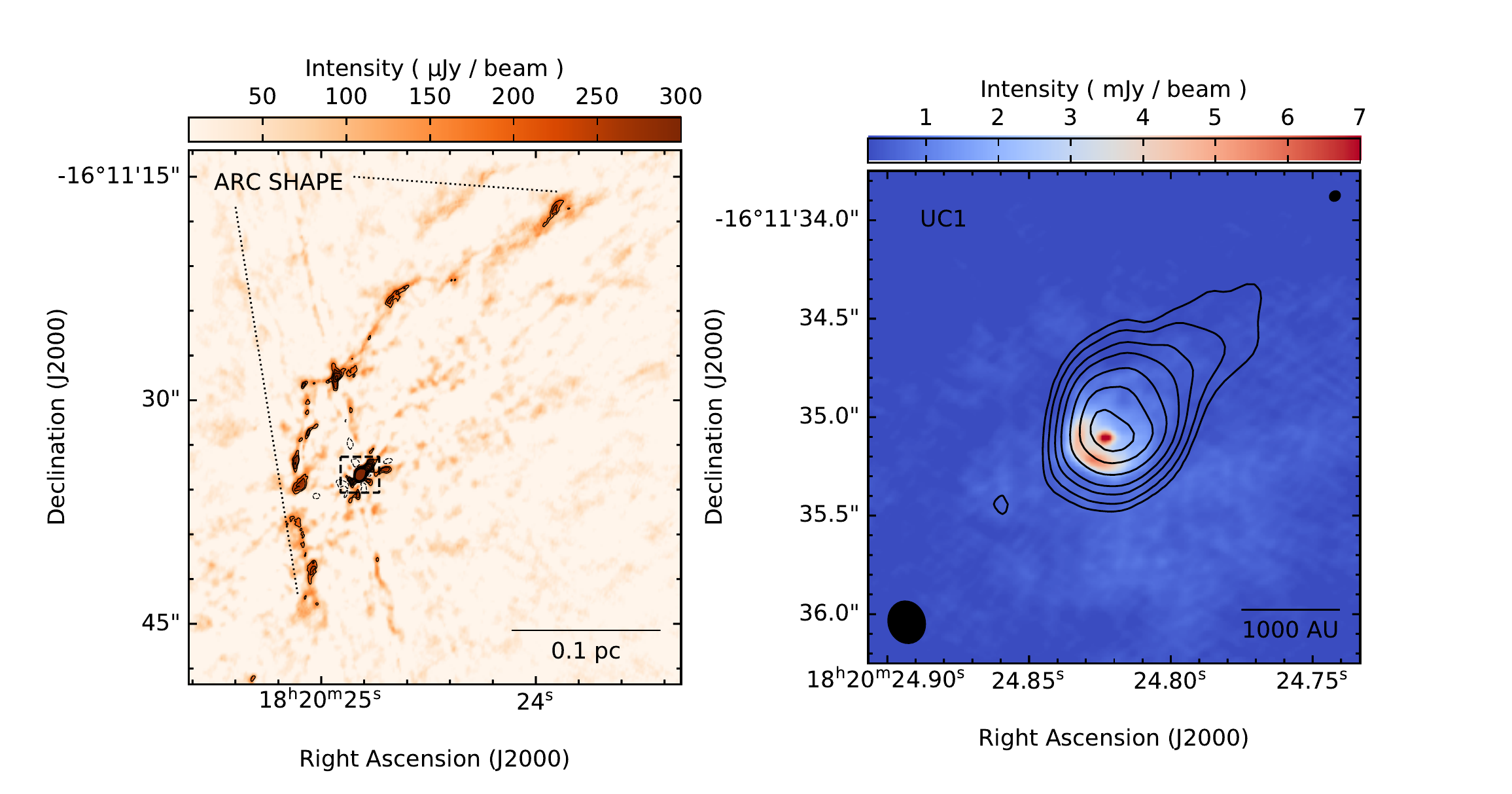}
 \caption{{\it Left Panel}: X band emission (color and contour scale) highlighting the arc shape structure (West part of the image). Contour levels are C times −2$^0$, 
 2$^{1/2}$, 
 2$^1$,
 2$^{3/2}$,
 2$^2$,
 2$^{5/2}$,
 2$^3$,
 2$^{7/2}$
 and 2$^4$,
 where C is given by 2.5 times the rms noise measured over the region of the map shown in the panel (70 $\mu$Jy beam$^{-1}$). The beam size is the same as shown in Fig. \ref{fig:bigim}. The dotted lines show the angular extend of the Arc shape. The dashed box represents the area shown enlarged in the right panel enclosing the UC1 source.   
 {\it Right Panel}: Detail of the X band emission (contours) of UC1 superimposed on the ALMA 240 GHz continuum image (color scale). Note the compact ALMA source that we identify as a disk associated with the ionizing star of the region. Contours are 3, 6, 12, 20, 35, 53 and 68 times the rms noise of the region of the map around UC1 (25 $\mu$Jy beam$^{-1}$). The corresponding beams are shown in the bottom left corner (VLA) and in the top right corner (ALMA).
 \label{fig:UC1}}
\end{figure*}

In Figure~\ref{fig:UC1} we focus on the X band emission at the western part of the cloud. The left panel of the figure shows the arc-shaped structure found previously by \citet{felli1980} and discussed by \citet{felli1984} and \citet{johnson1998}. This structure is identified as the ionizing front of the central HII region of M\,17. For the whole arc-like structure, we measure a flux density of 0.1 Jy and a bounding box of $\sim$ 5 $\times$ 30 arcsec. These values are significantly lower than those measured by \citet{felli1984} because our data suffer from severe spatial filtering.    

Near the focus of the arc morphology, \citet{felli1980} found a compact thermal radio source they named UC1 and which they categorized as an UCHII region. \citet{felli1984} determined a size ($\sim0.004$ pc) and spectral index ($\sim1$) that \citetalias{rodriguez2012} found more typical of an HCHII region.  Our measured flux density value at 10 GHz, $98 \pm 5$ mJy, is roughly consistent with the results of the spectral analysis of \citetalias{rodriguez2012}. However, \citetalias{rodriguez2012} showed that the spectral index derived for UC1 actually results from averaging a gradient along the cometary morphology highlighted in the right panel of Figure~\ref{fig:UC1} (countours). The gradient on the spectral index is likely a consequence of a gas density gradient in the source, being maximum at the tip of the cometary morphology at the south-east part of the source (i.e., implying emission measures of at least $\sim10^9$ cm$^{-6}$ pc).

We obtained archival ALMA band 6 data centered on UC1 in order to assess the scenario of RGM2012 described above (project 2015.1.01163). In agreement with this scheme, the mm emission integrated over the entire band (1.875 GHz) shows a bow-like structure oriented toward the SE, where high densities, either of dust or ionized material are expected to be found. In addition, a bright central source possibly associated with the stellar object of UC1 is also revealed in the mm map. The total integrated flux density of the mm emission centered at 228 GHz is about 125 mJy, 7 mJy of which come from this central source. The deconvolved size of the latter is $(30 \pm 10$ mas$) \times (22 \pm 17$ mas$)$ (P.A. of $70\arcdeg \pm 50\arcdeg$) that at the distance of M\,17 corresponds to $\sim60 \times 40$ AU. If this mm emission comes from a protostellar disk, its derived size is significantly lower than the sizes inferred for disks of other massive stars \citep[$\sim300$ AU and a few hundreds of mJy at similar wavelegths,][]{Lugo2004,girart2018}. A possibility is that the disk is being ablated by the energetic radiation of the protostellar object exciting UC1, although we cannot discard alternative scenarios involving other elements than the presence of a disk.

Exploring UC1 at IR wavelengths, \citet{nielbock2007} found hints of a disk-like structure in absorption across UC1, which was interpreted as a cool disk of 1000 AU in size. This size is more than an order of magnitude larger than the size of the central peak of the ALMA image. The fact that more extended material from this cool disk is not detected at the frequency of the ALMA observations suggests that it corresponds to a different structure, perhaps associated to the molecular part of the M\,17 region. The potential role of disks around massive stars, in particular  in their early stages (i.e., HCHII region) is presently unclear and their firm detection would have important implications on the paradigm of their formation and deserves further investigation.

\section{Conclusions} \label{6}

We have carried out deep X band observations with the VLA operating in A configuration. After self-calibrating the obtained data of M\,17 and extracting radio sources by applying the BLOBCAT software on the resulting map, we cataloged them and inspected their basic emission properties. We reached the following conclusions:

\begin{enumerate}

\item We found a total of 194 radio sources in the M\,17 region, 12 of them extended and 182 compact. This catalog constitutes the most complete catalog of radio sources known of M\,17. After Orion, M\,17 becomes the second richest region in radio source detection to our knowledge.

\item The inspection of the properties of our radio sources reveals a population (about 40\% of our catalog) of objects emitting non-thermal radiation. These are probably low-mass stars with coronal gyrosynchrotron emission, given their observed variability and polarization.

\item In X-ray emission, these non-thermal radio sources are underluminous with respect to the G\"udel Benz relation, but a reasonable correlation of their radio and X-ray emission is still found, provided that only sources with evident non thermal nature are selected from Orion and M\,17. This suggests that radio and X-ray emission are both caused  the same underlying process, presumably magnetic reconnection in the stellar coronae.

\item The luminosity function of the M\,17 radio sources shows a decreasing trend with luminosity (i.e. weak sources are more abundant than bright ones). The same trend is observed for the radio sources of the Orion Nebula Cluster and NGC\,6334D-F reported in other works, though there is an important lack of completeness for M\,17 and NGC\,6334D-F that prevents us from exploring this similarity for the weakest sources. In any case, M\,17 has about three more times sources at any given radio luminosity than the other two regions perhaps due to its higher intrinsic luminosity.

\item The spatial distribution of sources in M\,17 is centrally peaked towards the position of the most massive stars (e.g., CEN 1) and fades radially and smoothly toward the outskirts of the region, with a dispersion that is marginally higher for non thermal sources. The same analysis carried out for the radio sources in the ONC shows no segregation between thermal and non thermal sources. The radio cluster in M\,17 is about 5 times larger than that in the ONC.

\item  A previously known arc structure that has the HCHII region UC1 at its focus is also detected in our image. The cometary morphology of UC1 is highlighted in our data with a probable increase of density towards the ionization front traced by the arc. Exploring ALMA archival data at 1 mm, we found emission associated to the possible exciting source of UC1. If this emission comes from a protoplanetary disk, its size is considerably smaller than typical sizes inferred from disks around massive protostars. 

\end{enumerate}

\section*{Acknowledgements}

V.Y. acknowledges the financial support of
DAIP, UG, and MPIfR during the internship at the Max Planck Institute for Radioastronomy. V.Y. acknowledges the financial support of CONACyT, México. L.L. acknowledges the support of CONACyT-AEM grant 275201, CONACyT-CF grant 263356, and UNAM DGAPA/PAPIIT grants IN112417 and IN112820. This research has made use of the NASA/IPAC Infrared Science Archive, which is funded by the National Aeronautics and Space Administration and operated by the California Institute of Technology. Counterparts to radio sources were found by using SIMBAD database.

{\it Facilities:} This work makes use of the Spitzer, 2MASS, Gaia, Vary Large Array (VLA), and Atacama Large Millimiter Array (ALMA) telescopes.
The VLA observations presented here were part of NRAO
program 18A-002. The National Radio Astronomy Observatory
is a facility of the National Science Foundation operated under
cooperative agreement by Associated Universities, Inc. This paper makes use of the following ALMA data: ADS/JAO.ALMA\#2015.1.01163. ALMA is a partnership of ESO (representing its member states), NSF (USA) and NINS (Japan), together with NRC (Canada), MOST and ASIAA (Taiwan), and KASI (Republic of Korea), in cooperation with the Republic of Chile. The Joint ALMA Observatory is operated by ESO, AUI/NRAO and NAOJ.


\bibliography{M17_CRS}

\newpage

\appendix

\section{Catalog of the compact radio sources of M17}

\begin{center}
\footnotesize
\setlength\tabcolsep{2pt}
\begin{longtable}{c	c	c	c	c	c	c	c   c}
\caption{Parameters of the CRSs with their counterparts at other wavelengths.} \\
\hline
\hline
ID&$\alpha$ &$\delta$ & $S_{peak}$ & $S_{int}$&X-ray$^a$ & IR$^b$ & Optical$^c$ & Radio$^d$ \\  \textbf{\#}&(J2000)&(J2000)& $(\mu Jy~beam^{-1})$ & $(\mu Jy)$ \\ 
\hline
\endfirsthead
\multicolumn{9}{c}%
{\tablename\ \thetable\ -- \textit{Parameters of the CRSs with their counterparts at other wavelengths.}} \\ \hline
\hline
ID&$\alpha$ &$\delta$ & $S_{peak}$ & $S_{int}$&X-ray$^a$ & IR$^b$ & Optical$^c$ & Radio$^d$ \\ 
\textbf{\#}&(J2000)&(J2000) & $(\mu Jy~beam^{-1})$ & $(\mu Jy)$ \\ 
\hline
\endhead
\hline 
\multicolumn{9}{r}{\textit{}} \\
\endfoot 
\multicolumn{9}{l}{\textbf{}} \\
\endlastfoot

1 & 18 20  16.402 $\pm$ 0.002 & -16 11  18.81 $\pm$ 0.03 & 149 $\pm$ 30 & 179 $\pm$ 30 &  &  & & \\ 
2 & 18 20  16.405 $\pm$ 0.002 & -16 11  10.73 $\pm$ 0.03 & 143 $\pm$ 29 & 169 $\pm$ 30 &  &  &  &\\ 
3 & 18 20  17.569 $\pm$ 0.001 & -16 12  21.49 $\pm$ 0.02 & 188 $\pm$ 25 & 543 $\pm$ 118 &  &  &  &\\ 
4 & 18 20  18.968 $\pm$ 0.002 & -16 11  27.98 $\pm$ 0.03 &  88 $\pm$ 17 & 185 $\pm$ 19 &  &  &  &\\ 
5 & 18 20  19.417 $\pm$ 0.001 & -16 12  15.74 $\pm$ 0.01 & 549 $\pm$ 36 & 549 $\pm$ 36& \checkmark &  &  &\\ 
6 & 18 20  20.722 $\pm$ 0.001 & -16 08  27.79 $\pm$ 0.02 & 177 $\pm$ 29 & 241 $\pm$ 30 & \checkmark & \checkmark & \checkmark & \\ 
7 & 18 20  21.433 $\pm$ 0.001 & -16 12  06.75 $\pm$ 0.02 & 180 $\pm$ 17 & 192 $\pm$ 17& \checkmark & \checkmark & & 4 \\ 
8 & 18 20  21.636 $\pm$ 0.001 & -16 11  17.91 $\pm$ 0.01 & 286 $\pm$ 22 & 400 $\pm$ 25 &  & \checkmark & \checkmark & 5\\ 
9 & 18 20  21.736 $\pm$ 0.001 & -16 11  12.59 $\pm$ 0.01 & 222 $\pm$ 17 & 227 $\pm$ 16 &  &  &  & \\ 
10 & 18 20  21.849 $\pm$ 0.002 & -16 09  05.43 $\pm$ 0.03 &  72 $\pm$ 15 &  94 $\pm$ 15 &  &  &  &\\ 
11 & 18 20  21.921 $\pm$ 0.002 & -16 13  33.15 $\pm$ 0.03 & 201 $\pm$ 41 & 211 $\pm$ 41 &  &  &  &\\ 
12 & 18 20  22.339 $\pm$ 0.001 & -16 11  28.55 $\pm$ 0.02 &  97 $\pm$ 15 & 189 $\pm$ 17 &  &  &  &\\ 
13 & 18 20  22.358 $\pm$ 0.001 & -16 07  49.51 $\pm$ 0.01 & 187 $\pm$ 38 & 474 $\pm$ 44&  &  &  &\\ 
14 & 18 20  22.363 $\pm$ 0.001 & -16 12  04.87 $\pm$ 0.001 & 242 $\pm$ 19 & 242 $\pm$ 19 & \checkmark &  &  & 6 \\ 
15 & 18 20  22.468 $\pm$ 0.002 & -16 09  51.39  $\pm$ 0.03 &  55 $\pm$ 10 &  58 $\pm$ 10 & \checkmark &  &  &\\ 
16 & 18 20  22.852 $\pm$ 0.001 & -16 11  48.51 $\pm$ 0.02 & 171 $\pm$ 15 & 239 $\pm$ 17 & \checkmark &  &  &\\ 
17 & 18 20  23.082 $\pm$ 0.001 & -16 13  05.23 $\pm$ 0.02 & 244 $\pm$ 29 & 270 $\pm$ 29 &  &  &  &\\ 
18 & 18 20  23.107 $\pm$ 0.002 & -16 13  01.91 $\pm$ 0.03 & 123 $\pm$ 24 & 125 $\pm$ 24 & \checkmark &  &  &\\ 
19 & 18 20  23.157 $\pm$ 0.002 & -16 13  05.67 $\pm$ 0.03 & 134 $\pm$ 27 & 134 $\pm$ 27 & \checkmark &  &  &\\
20 & 18 20  23.181 $\pm$ 0.001 & -16 10  43.47 $\pm$ 0.02 & 119 $\pm$ 14 & 137 $\pm$15 &  &  &  &\\ 
21 & 18 20  23.763 $\pm$ 0.001 & -16 11  57.99 $\pm$ 0.02 & 112 $\pm$ 16 & 458 $\pm$ 27 &  &  &  &\\ 
22 & 18 20 24.821 $\pm$ 0.001 & -16 11  35.08 $\pm$ 0.01 & 18582 $\pm$ 1003 & 97836 $\pm$ 5092 & \checkmark &  \checkmark &  & UC1 \\
23 & 18 20  24.897 $\pm$ 0.001 & -16 10  42.56 $\pm$ 0.02 & 122 $\pm$ 15 & 122 $\pm$ 15 & \checkmark &  &  &\\ 
24 & 18 20  25.318 $\pm$ 0.001 & -16 11  48.64 $\pm$ 0.02 & 230 $\pm$ 29 & 892 $\pm$ 65 &  &  & & 7\\ 
25 & 18 20  25.496 $\pm$ 0.002 & -16 10  53.80 $\pm$ 0.03 &  90 $\pm$ 17 & 207 $\pm$ 20 & \checkmark & \checkmark & & 8\\ 
26 & 18 20  25.622 $\pm$ 0.002 & -16 09  37.04 $\pm$ 0.03 &  45 $\pm$  9 &  62 $\pm$  9 &  &  &  &\\ 
27 & 18 20  25.847 $\pm$ 0.001 & -16 08  32.16 $\pm$ 0.03 &  87 $\pm$ 13 &  93 $\pm$ 13 & \checkmark & \checkmark & \checkmark &\\ 
28 & 18 20  26.116 $\pm$ 0.002 & -16 09  13.92 $\pm$ 0.03 &  46 $\pm$  9 & 100 $\pm$ 10 &  &  &  &\\ 
29 & 18 20  26.216 $\pm$ 0.001 & -16 09  21.52 $\pm$ 0.02 &  70 $\pm$  10 &  85 $\pm$  10 &  & \checkmark & \checkmark  & \\ 
30 & 18 20  26.240 $\pm$ 0.001 & -16 11  15.84 $\pm$ 0.01 & 270 $\pm$ 21 & 345 $\pm$ 41 &  &  & & 10 \\ 
31 & 18 20  27.379 $\pm$ 0.001 & -16 11  38.56 $\pm$ 0.01 & 262 $\pm$ 18 & 565 $\pm$ 30 &  &  &  & 12\\ 
32 & 18 20  27.588 $\pm$ 0.001 & -16 11  33.40 $\pm$ 0.01 & 886 $\pm$ 49 & 886 $\pm$ 49 & \checkmark &  & \checkmark & \\ 
33 & 18 20  27.615 $\pm$ 0.001 & -16 10  43.32 $\pm$ 0.03 &  45 $\pm$  7 &  48 $\pm$  7 & \checkmark & \checkmark & \checkmark & \\ 
34 & 18 20  27.715 $\pm$ 0.001 & -16 09  51.12 $\pm$ 0.02 &  79 $\pm$  8 &  99 $\pm$  8 & \checkmark & \checkmark &  &\\
35 & 18 20  27.742 $\pm$ 0.002 & -16 14  16.76 $\pm$ 0.03 & 252 $\pm$ 49 & 340 $\pm$ 51 &  &  &  &\\ 
36 & 18 20  27.798 $\pm$ 0.001 & -16 08  57.76 $\pm$ 0.02 &  87 $\pm$ 11 &  87 $\pm$ 11 & \checkmark & \checkmark & \checkmark & \\ 
37 & 18 20  27.857 $\pm$ 0.001 & -16 09  55.64 $\pm$ 0.02 &  79 $\pm$  8 &  89 $\pm$  8 & \checkmark & \checkmark & \checkmark & 13\\ 
38 & 18 20  28.023 $\pm$ 0.001 & -16 11  26.96 $\pm$ 0.01 & 208 $\pm$ 14 & 274 $\pm$ 16 &  &  & & 14\\ 
39 & 18 20  28.034 $\pm$ 0.001 & -16 10  42.36 $\pm$ 0.02 &  79 $\pm$  7 &  80 $\pm$  7 &  &  &  &\\ 
40 & 18 20  28.151 $\pm$ 0.001 & -16 10  49.32 $\pm$ 0.03 &  41 $\pm$  7 &  72 $\pm$  7 & \checkmark & \checkmark & \checkmark & \\ 
41 & 18 20  28.154 $\pm$ 0.002 & -16 10  01.00 $\pm$ 0.03 &  31 $\pm$  6 &  31 $\pm$  6 &  &  &  &\\ 
42 & 18 20  28.206 $\pm$ 0.002 & -16 11  05.64 $\pm$ 0.03 &  37 $\pm$  7 &  37 $\pm$  7 &  &  &  &\\ 
43 & 18 20  28.231 $\pm$ 0.001 & -16 13  17.80 $\pm$ 0.01 & 1627 $\pm$ 89 & 1674 $\pm$ 85 & \checkmark & \checkmark & & 15\\
44 & 18 20  28.273 $\pm$ 0.001 & -16 10  42.36 $\pm$ 0.01 & 117 $\pm$  8 & 151 $\pm$  9 &  &  &  &\\ 
45 & 18 20  28.287 $\pm$ 0.001 & -16 11  30.56 $\pm$ 0.02 & 128 $\pm$ 13 & 172 $\pm$ 14 & \checkmark & \checkmark & & 16\\ 
46 & 18 20  28.367 $\pm$ 0.001 & -16 10  30.68 $\pm$ 0.02 &  43 $\pm$  6 &  43 $\pm$  6 & \checkmark & \checkmark & \checkmark &\\ 
47 & 18 20  28.384 $\pm$ 0.001 & -16 10  14.16 $\pm$ 0.01 & 164 $\pm$ 11 & 222 $\pm$ 13 & & & & 17\\ 
48 & 18 20  28.487 $\pm$ 0.001 & -16 10  39.56 $\pm$ 0.01 &  99 $\pm$  8 & 108 $\pm$ 8 &  &  &  &\\ 
49 & 18 20  28.503 $\pm$ 0.001 & -16 10  55.36 $\pm$ 0.02 &  58 $\pm$  7 & 226 $\pm$ 13 &  &  &  &\\ 
50 & 18 20  28.548 $\pm$ 0.001 & -16 10  26.84 $\pm$ 0.01 & 110 $\pm$  8 & 149 $\pm$  10 &  &  &  &\\ 
51 & 18 20  28.720 $\pm$ 0.001 & -16 10  22.16 $\pm$ 0.03 &  47 $\pm$  6 &  47 $\pm$  6 &  &  &  &\\ 
52 & 18 20  28.787 $\pm$ 0.002 & -16 11  51.92 $\pm$ 0.03 &  40 $\pm$  9 &  48 $\pm$  8 &  &  &  &\\ 
53 & 18 20  28.787 $\pm$ 0.001 & -16 10  49.32 $\pm$ 0.02 &  63 $\pm$  6 &  71 $\pm$  7 &  &  &  &\\ 
54 & 18 20  28.798 $\pm$ 0.001 & -16 11  03.68 $\pm$ 0.03 &  41 $\pm$  7 &  41 $\pm$  7 &  &  &  &\\ 
55 & 18 20  28.839 $\pm$ 0.001 & -16 10  56.08 $\pm$ 0.02 &  72 $\pm$  7 &  72 $\pm$  7 & \checkmark & \checkmark &  &\\ 
56 & 18 20  28.867 $\pm$ 0.001 & -16 11  10.08 $\pm$ 0.03 &  42 $\pm$  6 &  42 $\pm$  6 & \checkmark & \checkmark &  &\\ 
57 & 18 20  28.998 $\pm$ 0.001 & -16 10  33.44 $\pm$ 0.03 &  37 $\pm$  6 &  54 $\pm$  6 &  &  &  &\\ 
58 & 18 20  29.139 $\pm$ 0.001 & -16 12  09.84 $\pm$ 0.03 &  50 $\pm$  9 & 101 $\pm$ 10 &  &  &  &\\ 
59 & 18 20  29.167 $\pm$ 0.001 & -16 09  43.04 $\pm$ 0.01 & 124 $\pm$  9 & 156 $\pm$ 10 & \checkmark & \checkmark &  \checkmark &\\ 
60 & 18 20  29.192 $\pm$ 0.001 & -16 10  54.68 $\pm$ 0.01 &  28 $\pm$  6 &  28 $\pm$  6 &  &  &  &\\ 
61 & 18 20  29.217 $\pm$ 0.001 & -16 10  38.80 $\pm$ 0.02 &  49 $\pm$  6 &  61 $\pm$  6 &  &  &  &\\ 
62 & 18 20  29.231 $\pm$ 0.002 & -16 11  03.00 $\pm$ 0.03 &  34 $\pm$  6 &  51 $\pm$  6 &  &  &  &\\ 
63 & 18 20  29.267 $\pm$ 0.001 & -16 10  41.32 $\pm$ 0.01 &  83 $\pm$  7 &  83 $\pm$  7 & \checkmark & \checkmark & \checkmark &  \\ 
64 & 18 20  29.317 $\pm$ 0.001 & -16 10  47.96 $\pm$ 0.03 &  50 $\pm$  6 &  51 $\pm$  6 & \checkmark &  &  &\\ 
65 & 18 20  29.317 $\pm$ 0.001 & -16 09  50.60 $\pm$ 0.03 &  36 $\pm$  6 &  41 $\pm$  6 &  &  &  &\\ 
66 & 18 20  29.323 $\pm$ 0.002 & -16 10  18.08 $\pm$ 0.03 &  31 $\pm$  6 &  77 $\pm$  7 &  &  &  &\\ 
67 & 18 20  29.367 $\pm$ 0.001 & -16 11  08.88 $\pm$ 0.01 &  90 $\pm$  8 &  90 $\pm$  8 &  &  &  &\\ 
68 & 18 20  29.386 $\pm$ 0.002 & -16 09  43.00 $\pm$ 0.03 &  34 $\pm$  6 & 126 $\pm$  9 & \checkmark & \checkmark &  &\\ 
69 & 18 20  29.436 $\pm$ 0.001 & -16 10  49.84 $\pm$ 0.01 & 283 $\pm$ 16 & 656 $\pm$ 33 & \checkmark &  &  & 19\\ 
70 & 18 20  29.445 $\pm$ 0.002 & -16 09  16.64 $\pm$ 0.03 &  39 $\pm$  8 &  51 $\pm$  8 &  &  &  &\\ 
71 & 18 20  29.495 $\pm$ 0.002 & -16 08  04.00 $\pm$ 0.03 &  77 $\pm$ 16 & 101 $\pm$ 16 &  &  &  &\\ 
72 & 18 20  29.625 $\pm$ 0.002 & -16 11  41.56 $\pm$ 0.03 &  35 $\pm$  6 &  37 $\pm$  6 &  &  &  &\\ 
73 & 18 20  29.728 $\pm$ 0.001 & -16 10  59.92 $\pm$ 0.03 &  48 $\pm$  7 &  57 $\pm$  7 &  &  &  &\\ 
74 & 18 20  29.731 $\pm$ 0.001 & -16 11  04.48 $\pm$ 0.02 &  46 $\pm$  6 &  60 $\pm$  7 &  &  &  &\\ 
75 & 18 20  29.736 $\pm$ 0.001 & -16 10  53.40 $\pm$ 0.01 & 121 $\pm$  9 & 131 $\pm$  9 &  &  &  &\\ 
76 & 18 20  29.745 $\pm$ 0.001 & -16 09  55.64 $\pm$ 0.02 &  52 $\pm$  6 &  60 $\pm$  6 &  &  &  &\\ 
77 & 18 20  29.750 $\pm$ 0.002 & -16 11  25.76 $\pm$ 0.03 &  29 $\pm$  6 &  34 $\pm$  6 &  &  &  &\\ 
78 & 18 20  29.767 $\pm$ 0.001 & -16 10  44.72 $\pm$ 0.02 &  45 $\pm$  6 &  45 $\pm$  6 &  &  &  &\\ 
79 & 18 20  29.811 $\pm$ 0.001 & -16 10  45.56 $\pm$ 0.01 & 692 $\pm$ 38 & 891 $\pm$ 45 & \checkmark &  & & 20 (CEN 1b)\\ 
80 & 18 20  29.822 $\pm$ 0.001 & -16 11  11.80 $\pm$ 0.02 &  60 $\pm$  6 &  60 $\pm$  6 &  &  &  &\\ 
81 & 18 20  29.828 $\pm$ 0.001 & -16 10  48.32 $\pm$ 0.01 & 109 $\pm$  8 & 109 $\pm$  8 &  &  &  &\\ 
82 & 18 20  29.842 $\pm$ 0.001 & -16 10  40.72 $\pm$ 0.02 &  46 $\pm$  6 &  59 $\pm$  7 & \checkmark &  & & \\ 
83 & 18 20  29.842 $\pm$ 0.001 & -16 10  31.04 $\pm$ 0.01 & 137 $\pm$  9 & 198 $\pm$ 12 &  &  &  &\\ 
84 & 18 20  29.850 $\pm$ 0.001 & -16 10  41.20 $\pm$ 0.01 & 230 $\pm$ 14 & 230 $\pm$ 14 & \checkmark & & & \\ 
85 & 18 20  29.861 $\pm$ 0.001 & -16 10  44.72 $\pm$ 0.01 & 173 $\pm$ 11 & 177 $\pm$ 11 & &  &  & \\ 
86 & 18 20  29.897 $\pm$ 0.001 & -16 10  44.44 $\pm$ 0.01 & 566 $\pm$ 31 & 566 $\pm$ 31 & \checkmark & & & 21 (CEN 1a)\\ 
87 & 18 20  29.911 $\pm$ 0.002 & -16 10  35.08$\pm$ 0.03 &  30 $\pm$  6 &  30 $\pm$  6 & \checkmark &  &  &\\  
88 & 18 20  29.917 $\pm$ 0.001 & -16 10  33.60 $\pm$ 0.02 &  74 $\pm$  7 &  81 $\pm$  7 &  &  &  &\\ 
89 & 18 20  29.925 $\pm$ 0.001 & -16 09  14.64 $\pm$ 0.01 &  39 $\pm$  7 &  49 $\pm$  8 &  &  &  &\\ 
90 & 18 20  29.953 $\pm$ 0.001 & -16 10  50.80 $\pm$ 0.02 &  85 $\pm$  7 &  94 $\pm$  7 &  &  &  &\\ 
91 & 18 20  29.953 $\pm$ 0.002 & -16 10  33.76 $\pm$ 0.03 &  32 $\pm$  6 &  32 $\pm$  6 &  &  &  &\\ 
92 & 18 20  29.968 $\pm$ 0.001 & -16 11  15.36 $\pm$ 0.01 &  77 $\pm$  7 &  78 $\pm$  7 &  &  &  &\\ 
93 & 18 20  29.989 $\pm$ 0.001 & -16 10  44.68 $\pm$ 0.02 &  44 $\pm$  6 &  79 $\pm$  7 &  &  &  &\\ 
94 & 18 20  30.003 $\pm$ 0.001 & -16 10  35.28 $\pm$ 0.01 & 514 $\pm$ 28 & 627 $\pm$ 32 & \checkmark &  & & 22\\ 
95 & 18 20  30.031 $\pm$ 0.001 & -16 10  34.68 $\pm$ 0.02 &  79 $\pm$  7 &  79 $\pm$  7 &  &  &  &\\ 
96 & 18 20  30.042 $\pm$ 0.002 & -16 11  57.44 $\pm$ 0.03 &  35 $\pm$  7 &  35 $\pm$  7 & \checkmark & \checkmark & \checkmark &\\ 
97 & 18 20  30.044 $\pm$ 0.002 & -16 10  33.40 $\pm$ 0.03 &  32 $\pm$  6 &  34 $\pm$  6 &  &  &  &\\ 
98 & 18 20  30.094 $\pm$ 0.001 & -16 10  41.44 $\pm$ 0.02 &  44 $\pm$  6 &  54 $\pm$  6 & \checkmark &  &  &\\ 
99 & 18 20  30.133 $\pm$ 0.001 & -16 10  39.48 $\pm$ 0.01 & 307 $\pm$ 17 & 404 $\pm$ 21 &  &  &  & 23\\  
100 & 18 20  30.178 $\pm$ 0.001 & -16 11  20.08 $\pm$ 0.02 &  58 $\pm$  6 &  58 $\pm$  6 &  &  &  & \\ 
101 & 18 20  30.194 $\pm$ 0.002 & -16 10  50.48 $\pm$ 0.03 &  34 $\pm$  6 &  34 $\pm$  6 & \checkmark &  & & \\ 
102 & 18 20  30.208 $\pm$ 0.001 & -16 10  12.04 $\pm$ 0.01 &  85 $\pm$  7 &  89 $\pm$  7 &  &  &  &\\ 
103 & 18 20  30.278 $\pm$ 0.001 & -16 11  31.48 $\pm$ 0.03 &  37 $\pm$  6 &  49 $\pm$  6 & \checkmark & \checkmark &\checkmark  & \\ 
104 & 18 20  30.286 $\pm$ 0.001 & -16 10  39.52 $\pm$ 0.03 &  34 $\pm$  6 &  48 $\pm$  6 &  &  &  &\\ 
105 & 18 20  30.325 $\pm$ 0.001 & -16 10  50.68 $\pm$ 0.01 & 186 $\pm$ 11 & 198 $\pm$ 11 &  &  &  & 24\\ 
106 & 18 20  30.380 $\pm$ 0.002 & -16 11  44.20 $\pm$ 0.03 &  28 $\pm$  6 &  32 $\pm$  6 &  &  &  & \\ 
107 & 18 20  30.394 $\pm$ 0.001 & -16 10  52.64 $\pm$ 0.03 &  39 $\pm$  6 &  48 $\pm$  6 &  &  &  &\\ 
108 & 18 20  30.414 $\pm$ 0.001 & -16 11  03.64 $\pm$ 0.01 & 179 $\pm$ 11 & 218 $\pm$ 13 &  &  &  &26\\ 
109 & 18 20  30.436 $\pm$ 0.001 & -16 11  01.56 $\pm$ 0.02 &  91 $\pm$  8 & 105 $\pm$  9 &  &  &  &\\ 
110 & 18 20  30.444 $\pm$ 0.001 & -16 10  53.12 $\pm$ 0.01 & 152 $\pm$  10 & 153 $\pm$  10 & \checkmark & \checkmark & \checkmark & 27\\ 
111 & 18 20  30.450 $\pm$ 0.001 & -16 10  10.08 $\pm$ 0.03 &  35 $\pm$  5 &  78 $\pm$  6 & \checkmark & \checkmark & \checkmark &  \\ 
112 & 18 20  30.530 $\pm$ 0.002 & -16 11  51.40 $\pm$ 0.03 &  32 $\pm$  6 &  42 $\pm$  6 &  &  &  &\\ 
113 & 18 20  30.541 $\pm$ 0.001 & -16 10  39.56 $\pm$ 0.02 &  46 $\pm$  6 &  53 $\pm$  6 &  &  &  &\\ 
114 & 18 20  30.575 $\pm$ 0.001 & -16 10  30.96 $\pm$ 0.03 &  36 $\pm$  6 &  47 $\pm$  6 &  &  &  &\\ 
115 & 18 20  30.578 $\pm$ 0.001 & -16 11  04.16 $\pm$ 0.01 & 606 $\pm$ 33 & 670 $\pm$ 35 &  & \checkmark &  & 28\\ 
116 & 18 20  30.633 $\pm$ 0.002 & -16 10  28.52 $\pm$ 0.03 &  27 $\pm$  5 &  27 $\pm$  5 & \checkmark & \checkmark &  & 29\\ 
117 & 18 20  30.664 $\pm$ 0.001 & -16 10  44.24 $\pm$ 0.03 &  35 $\pm$  6 &  46$\pm$  6 &  &  & & \\ 
118 & 18 20  30.678 $\pm$ 0.001 & -16 10  50.12 $\pm$ 0.02 &  61 $\pm$  6 &  73 $\pm$  7 &  &  &  &\\ 
119 & 18 20  30.711 $\pm$ 0.001 & -16 10  31.96 $\pm$ 0.02 &  45 $\pm$  6 &  45 $\pm$  6 &  &  &  &\\ 
120 & 18 20  30.777 $\pm$ 0.001 & -16 10  59.36$\pm$ 0.01 & 585 $\pm$ 32 & 826 $\pm$ 42 & \checkmark &  & & 30\\ 
121 & 18 20  30.780 $\pm$ 0.001 & -16 11  00.24 $\pm$ 0.01 &  29 $\pm$  6 &  75 $\pm$  7 &  &  &  &\\ 
122 & 18 20  30.811 $\pm$ 0.001 & -16 10  41.60 $\pm$ 0.02 &  61 $\pm$  6 &  61 $\pm$  6 & \checkmark &  &  &\\ 
123 & 18 20  30.850 $\pm$ 0.001 & -16 10  48.76 $\pm$ 0.02 &  74 $\pm$  7 & 166 $\pm$  10 &  &  &  &\\ 

124 & 18 20  30.877 $\pm$ 0.002 & -16 09  05.04 $\pm$ 0.03 &  40 $\pm$  8 &  44 $\pm$  8 & \checkmark &  \checkmark & \checkmark &  \\ 
125 & 18 20  30.894 $\pm$ 0.001 & -16 09  54.80 $\pm$ 0.02 &  60 $\pm$  6 &  84 $\pm$  7 & \checkmark & \checkmark & \checkmark &  \\ 
126 & 18 20  30.908 $\pm$ 0.001 & -16 11  35.96 $\pm$ 0.03 &  34 $\pm$  6 &  38 $\pm$ 6 & \checkmark & \checkmark & \checkmark &  \\ 
127 & 18 20  30.911 $\pm$ 0.001 & -16 10  39.48 $\pm$ 0.01 &  98 $\pm$  7 & 135 $\pm$  9 & \checkmark & \checkmark &  &\\
128 & 18 20  31.111 $\pm$ 0.001 & -16 09  29.84 $\pm$ 0.01 & 147 $\pm$ 10 & 147 $\pm$ 10 & \checkmark & \checkmark & \checkmark & 31 \\ 
129 & 18 20  31.305 $\pm$ 0.001 & -16 10  50.96 $\pm$ 0.02 &  39 $\pm$  5 &  39 $\pm$  5 & \checkmark &  &  &\\ 
130 & 18 20  31.336 $\pm$ 0.001 & -16 10  52.80 $\pm$ 0.02 &  38 $\pm$  5 &  41 $\pm$  5 &  &  &  &\\ 
131 & 18 20  31.355 $\pm$ 0.001 & -16 10  57.84 $\pm$ 0.01 & 125 $\pm$  8 & 130 $\pm$  8 &  &  &  &\\ 
132 & 18 20  31.424 $\pm$ 0.002 & -16 11  03.48 $\pm$ 0.03 &  28 $\pm$  5 &  28 $\pm$  6 &  &  &  &\\ 
133 & 18 20  31.433 $\pm$ 0.001 & -16 10  31.00 $\pm$ 0.01 &  75 $\pm$  6 &  93 $\pm$  7 &  &  &  &\\ 
134 & 18 20  31.469 $\pm$ 0.001 & -16 09  13.08 $\pm$ 0.03 &  47 $\pm$  8 &  74 $\pm$  8 &  &  &  &\\ 
135 & 18 20  31.477 $\pm$ 0.002 & -16 09  48.76 $\pm$ 0.03 &  28 $\pm$  6 &  32 $\pm$  6 &  &  &  &\\ 
136 & 18 20  31.516 $\pm$ 0.001 & -16 09  10.40 $\pm$ 0.02 &  65 $\pm$  8 &  78 $\pm$  8 & \checkmark &  &\checkmark &  \\ 
137 & 18 20  31.527 $\pm$ 0.002 & -16 10  32.60 $\pm$ 0.03 &  28 $\pm$  5 &  39 $\pm$  6 & \checkmark & \checkmark &  & 32\\ 
138 & 18 20  31.602 $\pm$ 0.001 & -16 11  05.88 $\pm$ 0.02 &  58 $\pm$  6 &  58 $\pm$  6 &  &  &  &\\ 
139 & 18 20  31.641 $\pm$ 0.002 & -16 10  36.72 $\pm$ 0.03 &  28 $\pm$  6 &  44 $\pm$  6 &  &  &  &\\ 
140 & 18 20  31.716 $\pm$ 0.001 & -16 11  06.80 $\pm$ 0.03 &  39 $\pm$  6 &  39 $\pm$  6 &  &  &  &\\ 
141 & 18 20  31.883 $\pm$ 0.001 & -16 11  24.04 $\pm$ 0.03 &  34 $\pm$  6 &  42 $\pm$  6 & \checkmark & \checkmark &  &\\ 
142 & 18 20  31.977 $\pm$ 0.001 & -16 09  25.16 $\pm$ 0.02 &  55 $\pm$  7 &  55 $\pm$  7 & \checkmark & \checkmark &  &\\ 
143 & 18 20  31.999 $\pm$ 0.002 & -16 11  27.00 $\pm$ 0.03 &  30 $\pm$  6 &  32 $\pm$  6 & \checkmark &  &  &\\ 
144 & 18 20  32.002 $\pm$ 0.002 & -16 09  40.44 $\pm$ 0.03 &  31 $\pm$  6 &  41 $\pm$  6 &  &  &  &\\ 
145 & 18 20  32.324 $\pm$ 0.001 & -16 10  48.92 $\pm$ 0.01 &  92 $\pm$  7 & 102 $\pm$  7 &  &  &  &\\ 
146 & 18 20  32.521 $\pm$ 0.002 & -16 11  11.04 $\pm$ 0.03 &  32 $\pm$  6 &  32 $\pm$  6 &  &  &  &\\ 
147 & 18 20  32.549 $\pm$ 0.002 & -16 10  06.20 $\pm$ 0.03 &  27 $\pm$  5 &  38 $\pm$  5 &  &  &  &\\ 
148 & 18 20  32.688 $\pm$ 0.002 & -16 10  20.16 $\pm$ 0.03 &  26 $\pm$  5 &  32 $\pm$  5 &  &  &  &\\ 
149 & 18 20  32.932 $\pm$ 0.001 & -16 10  35.68 $\pm$ 0.02 &  46 $\pm$  6 &  46 $\pm$  6 &  &  &  &\\ 
150 & 18 20  32.979 $\pm$ 0.001 & -16 11  15.56 $\pm$ 0.01 & 107 $\pm$  8 & 117 $\pm$  8 &  &  &  & 33\\ 
151 & 18 20  33.060 $\pm$ 0.001 & -16 11  21.60 $\pm$  0.01 & 642 $\pm$ 35 & 642 $\pm$ 35 & \checkmark & \checkmark &  & 34\\ 
152 & 18 20  33.093 $\pm$ 0.001 & -16 09  40.80 $\pm$ 0.01 & 153 $\pm$ 10 & 739 $\pm$ 37 &  &  &  & 35\\ 
153 & 18 20  33.199 $\pm$ 0.002 & -16 11  20.84 $\pm$ 0.03 &  29 $\pm$  6 &  29 $\pm$  6 &  &  & & \\ 
154 & 18 20  33.247 $\pm$ 0.001 & -16 13  22.92 $\pm$ 0.03 & 109 $\pm$ 18 & 109 $\pm$ 18 &  &  &  &\\ 
155 & 18 20  33.324 $\pm$ 0.002 & -16 11  11.00 $\pm$ 0.03 &  25 $\pm$  5 &  25 $\pm$  5 &  &  &  &\\ 
156 & 18 20  33.398 $\pm$ 0.002 & -16 10  18.44 $\pm$ 0.03  &  29 $\pm$  5 &  34 $\pm$  6 &  &  &  &\\ 
157 & 18 20  33.404 $\pm$ 0.001 & -16 09  44.40 $\pm$ 0.02 &  65 $\pm$  7 &  77 $\pm$  7 &  &  &  &\\ 
158 & 18 20  33.438 $\pm$ 0.002 & -16 10  55.32 $\pm$ 0.03 &  29 $\pm$  6 &  72 $\pm$  7 &  &  &  &\\ 
159 & 18 20  33.504 $\pm$ 0.001 & -16 10  44.60 $\pm$ 0.01 & 235 $\pm$ 14 & 352 $\pm$ 18 &  &  &  &36\\ 
160 & 18 20  33.610 $\pm$ 0.001 & -16 10  21.68 $\pm$ 0.01 & 115 $\pm$  8 & 136 $\pm$  9 &  &  &  &37\\ 
161 & 18 20  33.768 $\pm$ 0.001 & -16 10  46.44 $\pm$ 0.01 & 127 $\pm$  9 & 176 $\pm$ 10 &  &  &  &38\\ 
162 & 18 20  34.059 $\pm$ 0.001 & -16 10  28.88 $\pm$ 0.02 &  50 $\pm$  6 &  65 $\pm$  7 &  &  &  &\\ 
163 & 18 20  34.079 $\pm$ 0.001 & -16 10  44.52 $\pm$ 0.01 &  83 $\pm$  7 &  91 $\pm$  7 & \checkmark & \checkmark & \checkmark &   \\ 
164 & 18 20  34.209 $\pm$ 0.002 & -16 10  21.36 $\pm$ 0.03 &  29 $\pm$  6 &  53 $\pm$  6 &  &  &  &\\ 
165 & 18 20  34.576 $\pm$ 0.001 & -16 10  23.44 $\pm$ 0.01 &  85 $\pm$  7 &  93 $\pm$  7 &  &  &  &\\ 
166 & 18 20  34.673 $\pm$ 0.001 & -16 11  23.40 $\pm$ 0.02 &  67 $\pm$  7 &  79 $\pm$  7 &  &  &  &\\ 
167 & 18 20  34.888 $\pm$ 0.002 & -16 13  47.60 $\pm$ 0.03 & 160 $\pm$ 32 & 213 $\pm$ 33 &  &  &  &\\ 
168 & 18 20  35.048 $\pm$ 0.001 & -16 09  48.28 $\pm$ 0.02 &  79 $\pm$  8 & 135 $\pm$  10 &  &  &  &\\
169 & 18 20  35.197 $\pm$ 0.001 & -16 09  49.64 $\pm$ 0.02 &  72 $\pm$  8 & 121 $\pm$  9 &  &  &  &\\ 
170 & 18 20  35.479 $\pm$ 0.001 & -16 11  11.84 $\pm$ 0.02 &  75 $\pm$  7 &  77 $\pm$  7 & \checkmark & \checkmark & \checkmark &\\ 
171 & 18 20  35.858 $\pm$ 0.002 & -16 10  18.95 $\pm$ 0.03 &  37 $\pm$  7 &  69 $\pm$  8 &  &  &  &\\ 
172 & 18 20  36.238 $\pm$ 0.002 & -16 08  57.43 $\pm$ 0.03 &  54 $\pm$ 11 &  54 $\pm$ 11 &  &  &  &\\ 
173 & 18 20  36.449 $\pm$ 0.002 & -16 08  17.03 $\pm$ 0.03 &  87 $\pm$ 17 & 167 $\pm$ 19 &  &  &  &\\ 
174 & 18 20  36.879 $\pm$ 0.001 & -16 12  14.91 $\pm$ 0.02 &  99 $\pm$ 11 & 232 $\pm$ 15 &  &  &  &\\ 
175 & 18 20  37.762 $\pm$ 0.002 & -16 08  43.03 $\pm$ 0.03 &  75 $\pm$ 15 &  83 $\pm$ 15 &  &  &  &\\ 
176 & 18 20  37.803 $\pm$ 0.002 & -16 11  53.83 $\pm$ 0.03 &  53 $\pm$  10 &  58 $\pm$  10 &  &  &  &\\ 
177 & 18 20  37.895 $\pm$ 0.002 & -16 12  19.99 $\pm$ 0.03 &  57 $\pm$ 12 &  58 $\pm$ 12 &  &  &  &\\ 
178 & 18 20  39.819 $\pm$ 0.002 & -16 11  37.47 $\pm$ 0.03 &  57 $\pm$ 12 &  57 $\pm$ 12 &  &  &  &\\ 
179 & 18 20  40.895 $\pm$ 0.002 & -16 10  08.46 $\pm$ 0.03 &  67 $\pm$ 14 &  67 $\pm$ 14 &  &  &  &\\ 
180 & 18 20  40.984 $\pm$ 0.002 & -16 10  08.30 $\pm$ 0.03 &  71 $\pm$ 14 &  72 $\pm$ 14 &  &  &  &\\ 
181 & 18 20  43.625 $\pm$ 0.002 & -16 10  38.57 $\pm$ 0.03 & 133 $\pm$ 26 & 178 $\pm$ 27 &  &  &  &\\ 
182 & 18 20  44.343 $\pm$ 0.002 & -16 10  09.77 $\pm$ 0.03 & 207 $\pm$ 38 & 273 $\pm$ 39 &  &  &  &\\ 
\hline
\label{tab:CRS}
\end{longtable}
\begin{tablenotes}
      \item $^a$ {Chandra X-ray Observatory, \citep{broos2007}}
      \item $^b$ {Two Micron All-Sky Survey, Spitzer Space Telescope Galactic Legacy infrared Mid-Plane survey. IR counterparts are reported when the sources are detected at least at one of the available bands.}
      \item $^c$ {Gaia Data Release 2.} 
      \item $^d$ {Catalog number in \citetalias{rodriguez2012}}.
\end{tablenotes}
\end{center}

\end{document}